\documentclass[floatfix, showpacs,showkeys,aps,prd,twocolumn]{revtex4-1}

\usepackage{lineno}
\usepackage{amsmath}
\usepackage{amssymb}

\usepackage{tikz}
\usepackage{lipsum,adjustbox}
\usepackage{graphicx}
\usepackage[caption = false]{subfig}
\usepackage{dcolumn}
\usepackage{bm}
\usepackage{gensymb}

\usepackage{tabstackengine}
\usepackage[inline]{enumitem}

\setstackEOL{\cr}
\setstackgap{L}{\normalbaselineskip}

\let\oldequation\equation
\let\oldendequation\endequation

\renewenvironment{equation}
  {\linenomathNonumbers\oldequation}
  {\oldendequation\endlinenomath}

\newcolumntype{L}[1]{>{\raggedright\arraybackslash}p{#1}}
\newcolumntype{C}[1]{>{\centering\arraybackslash}p{#1}}
\newcolumntype{R}[1]{>{\raggedleft\arraybackslash}p{#1}}

\usepackage{hyperref}
\hypersetup{
 pdftitle={},
 pdfauthor={},
 pdfsubject={},
 pdfkeywords={},
 pdfstartview={},
 bookmarksopen=true, breaklinks=true, debug=true,
 colorlinks=true, linkcolor=blue, citecolor=red, urlcolor=blue,
 hyperfigures=true
}

\def \b{{\cal B}}
\usepackage{relsize}
\def \babar{\mbox{\slshape B\kern-0.1em{\smaller a}\kern-0.1em
    B\kern-0.1em{\smaller a\kern-0.1em r}}}

\usepackage{diagbox}
\begin{document}

\title{\boldmath Updated measurement of the branching fraction of $D_s^+\to \tau^+\nu_\tau$ via $\tau^+\to\pi^+\bar{\nu}_\tau$}

\author{
\begin{small}
\begin{center}
M.~Ablikim$^{1}$, M.~N.~Achasov$^{13,b}$, P.~Adlarson$^{75}$, R.~Aliberti$^{36}$, A.~Amoroso$^{74A,74C}$, M.~R.~An$^{40}$, Q.~An$^{71,58}$, Y.~Bai$^{57}$, O.~Bakina$^{37}$, I.~Balossino$^{30A}$
% ...
\\
\vspace{0.2cm}
(BESIII Collaboration)\\
\vspace{0.2cm} {\it
$^{1}$ Institute of High Energy Physics, Beijing 100049, People's Republic of China\\
% ...
$^{13}$ G.I. Budker Institute of Nuclear Physics SB RAS (BINP), Novosibirsk 630090, Russia\\
% ...
$^{30}$ INFN Sezione di Ferrara, (A)INFN Sezione di Ferrara, I-44122, Ferrara, Italy; (B)University of Ferrara,  I-44122, Ferrara, Italy\\
% ...
$^{36}$ Johannes Gutenberg University of Mainz, Johann-Joachim-Becher-Weg 45, D-55099 Mainz, Germany\\
$^{37}$ Joint Institute for Nuclear Research, 141980 Dubna, Moscow region, Russia\\
% ...
$^{40}$ Liaoning Normal University, Dalian 116029, People's Republic of China\\
% ...
$^{57}$ Southeast University, Nanjing 211100, People's Republic of China\\
$^{58}$ State Key Laboratory of Particle Detection and Electronics, Beijing 100049, Hefei 230026, People's Republic of China\\
% ...
$^{71}$ University of Science and Technology of China, Hefei 230026, People's Republic of China\\
% ...
$^{74}$ University of Turin and INFN, (A)University of Turin, I-10125, Turin, Italy; (B)University of Eastern Piedmont, I-15121, Alessandria, Italy; (C)INFN, I-10125, Turin, Italy\\
$^{75}$ Uppsala University, Box 516, SE-75120 Uppsala, Sweden\\
% ...
$^{b}$ Also at the Novosibirsk State University, Novosibirsk, 630090, Russia\\
% ...
}\end{center}
\vspace{0.4cm}
\end{small}}

\date{\today}

\begin{abstract}
This paper reports the study of $D_s^+\to \tau^+\nu$ via $\tau^+\to\pi^+\bar{\nu}_\tau$ using a boosted decision tree method, with $7.33$ fb$^{-1}$ of $e^{+}e^{-}$ collision data collected by the BESIII detector at center-of-mass energies between $4.128$ and $4.226$~GeV. The branching fraction of $D_s^+\to \tau^+\nu_\tau$ is determined to be $(5.44\pm0.17_{\rm stat}\pm0.13_{\rm syst})\%$. The product of the $D_s^+$ decay constant $f_{D_s^+}$ and the CKM matrix element $|V_{cs}|$ is $f_{D_s^+}|V_{cs}| = (248.3\pm3.9_{\rm stat}\pm3.1_{\rm syst}\pm1.0_{\rm input})~\mathrm{MeV}$. Combining with the $|V_{cs}|$ value obtained from the Standard Model global fit or the $f_{D_s^+}$ from the lattice quantum chromodynamics, we determine $|V_{cs}| =  0.993\pm0.015_{\rm stat}\pm0.012_{\rm syst}\pm0.004_{\rm input}$ and $f_{D_s^+} = (255.0\pm4.0_{\rm stat}\pm3.2_{\rm syst}\pm1.0_{\rm input})~\text{MeV}$.
The first uncertainty is statistical, the second one is systematic and the third one is due to the input parameters, mainly the lifetime of $D_s^+$.
All results obtained in this work supersede the BESIII previous results based on 6.32 fb$^{-1}$ of $e^+e^-$ collision data taken at center-of-mass energies between $4.178$ and $4.226$~GeV.
\end{abstract}

\maketitle

\section{INTRODUCTION}\label{sec:intro}
In the leptonic decay $D_s^+ \rightarrow \ell^+ \nu_\ell$, the charm quark ($c$) and anti-strange quark ($\bar{s}$) annihilate through a virtual $W$ boson to a charged and neutral lepton pair. According to the Standard Model (SM) and ignoring radiative corrections, the partial decay width of $D_s^+ \rightarrow \ell^+ \nu_\ell$ can be written as~\cite{Li:2021iwf} 
\begin{equation}\label{eq:decayrate}
  \begin{aligned}
    \Gamma_{D_s^+\to\ell^+\nu_\ell} 
    &=\frac{{\b}_{D^+_s\to \ell^+\nu_\ell}}{\tau_{D^+_s}}\\
    &=\frac{G^2_F}{8\pi}f^2_{D_s^+}|V_{cs}|^2m^2_\ell m_{D_s^+}\left(1-\frac{m^2_\ell}{m^2_{D_s^+}}\right)^2,\\
  \end{aligned}
\end{equation}
\noindent where ${\b}_{D^+_s\to \ell^+\nu_\ell}$ is the branching fraction of $D^+_s\to \ell^+\nu_\ell$, $\tau_{D^+_s}$ is the lifetime of $D^+_s$, $m_{D_s^+}$ is the mass of $D_s^+$, $m_\ell$ is the mass of the lepton $\ell^+$, $G_F$ is the Fermi coupling constant, $f_{D^+_s}$ is the $D^+_s$ decay constant describing strong effect between quarks, $|V_{cs}|$ is the $c\to s$ Cabibbo-Kobayashi-Maskawa (CKM) matrix element describing weak effect between quarks. Measurement of the branching fraction of $D_s^+\to \ell^+\nu_\ell$ can help us to determine $f_{D_s^+}$ when taking the $|V_{cs}|$ from the SM global fit as input, thereby testing various theoretical predictions, especially those from lattice quantum chromodynamics (LQCD)~\cite{FNAL/MILC17, ETM14E, chiQCD20A, RBC/UKQCD18A, HPQCD12A, PACS-CS11, Balasubramamian19, Blossier18, TWQCD14, ALPHA13B}. Conversely, one can determine $|V_{cs}|$ by taking the LQCD calculation of $f_{D_s^+}$ as input, thereby providing a stricter test of the CKM matrix unitarity.

In addition, the ratio of the $D_s^+\to \tau^+\nu_\tau$ and $D_s^+\to \mu^+\nu_\mu$ partial decay widths is defined as\begin{equation}\label{eq:theR}
  R = \frac{\Gamma_{D_s^+\to\tau^+\nu_\tau}}{\Gamma_{D_s^+\to\mu^+\nu_\mu}} = \frac{m^2_\tau(1-\frac{m^2_\tau}{m^2_{D_s^+}})^2}{m^2_\mu(1-\frac{m^2_\mu}{m^2_{D_s^+}})^2},
\end{equation}
\noindent where $f_{D_s^+}$ and $|V_{cs}|$ have been canceled. The SM calculation gives a very precise prediction of $R = 9.75\pm0.01$. Any significant deviation from this value would imply new physics beyond the SM.

The measurements of the branching fraction of $D_s^+\to\tau^+\nu_\tau$ have been reported by CLEO~\cite{CLEOel,CLEOpi,CLEOro}, BaBar~\cite{BaBarlnu}, Belle~\cite{Bellelnu} and BESIII~\cite{hajime, BES3taupipi0,BES3taumu,BES3taue,kebaiqian}. Among them, BESIII reported the measurements of the branching fraction of $D_s^+\to\tau^+\nu_\tau$ via $\tau^+\to\pi^+\bar{\nu}_\tau$~\cite{hajime}, $\tau^+\to\pi^+\pi^0\bar{\nu}_\tau$~\cite{BES3taupipi0}, $\tau^+\to e^+\bar{\nu}_\tau\nu_e$~\cite{BES3taue}, and $\tau^+\to \mu^+\bar{\nu}_\tau\nu_\mu$~\cite{BES3taumu} using $6.32$ fb$^{-1}$ of $e^+e^-$ collision data collected at center-of-mass energies ($E_{\rm cm}$) between 4.178 and 4.226 GeV, as well as the measurement via $\tau^+\to\pi^+\bar{\nu}_\tau$ using 0.48 fb$^{-1}$ of $e^+e^-$ collision data at $E_{\rm cm}=4.008$ GeV~\cite{BES34009}.
This paper presents an updated measurement of the branching fraction of $D_s^+\to \tau^+\nu_\tau$ via $\tau^+\to \pi^+\bar{\nu}_\tau$ with the boosted decision tree (BDT)~\cite{tmva4} method where the BDT output score is used to extract the signal yield. This analysis is based on 7.33 fb$^{-1}$ of $e^+e^-$ collision data taken at $E_{\text{cm}}$ = $4.128$~GeV, $4.157$~GeV, $4.178$~GeV, $4.189$~GeV, $4.199$~GeV, $4.209$~GeV, $4.219$~GeV, and $4.226$~GeV~\cite{ecm4230}. The integrated luminosities for these subsamples are $0.402$~fb$^{-1}$, $0.409$~fb$^{-1}$, $3.189$~fb$^{-1}$, $0.570$~fb$^{-1}$, $0.526$~fb$^{-1}$, $0.572$~fb$^{-1}$, $0.569$~fb$^{-1}$, and $1.092$~fb$^{-1}$~\cite{lumi4230}, respectively, with an uncertainty of 1\% dominated by systematic uncertainty. Compared to Ref.~ \cite{hajime}, the data sets at the former two energy points are newly added, the $D^*_s\to \pi^0D_s$ chain is used, and the range of the missing mass square of the missing neutrinos of the signal candidates is extended from $[-0.2,0.2]$ GeV$^2/c^4$ to $[-0.2,0.6]$ GeV$^2/c^4$. The reported results in this work supersede the previous results reported in Ref.~\cite{hajime}. Throughout this paper, charge conjugate modes are always implied.

\section{BESIII EXPERIMENT AND DATA SETS}
The BESIII detector~\cite{2009MAblikimDet} records symmetric $e^+e^-$ collisions provided by the BEPCII storage ring~\cite{bepcii} in the center-of-mass energy range from 2.0~GeV to 4.95~GeV, with a peak luminosity of $1 \times 10^{33}\;\text{cm}^{-2}\text{s}^{-1}$ achieved at $\sqrt{s} = 3.77\;\text{GeV}$. BESIII has collected large data samples in this energy region~\cite{2009MAblikimDet}. The cylindrical core of the BESIII detector covers 93\% of the full solid angle and consists of a helium-based multilayer drift chamber~(MDC), a plastic scintillator time-of-flight system~(TOF), and a CsI(Tl) electromagnetic calorimeter~(EMC), which are all enclosed in a superconducting solenoidal magnet providing a 1.0~T magnetic field~\cite{detvis}. The solenoid is supported by an octagonal flux-return yoke with resistive plate counter muon identification modules interleaved with steel. The charged particle momentum resolution at $1~{\rm GeV}/c$ is $0.5\%$, and the specific ionization energy loss ${\rm d}E/{\rm d}x$ resolution is $6\%$ for electrons from Bhabha scattering. The EMC measures photon energies with a resolution of $2.5\%$ ($5\%$) at $1$~GeV in the barrel (end-cap) region. The time resolution in the TOF barrel region is 68~ps. The end-cap TOF system was upgraded in 2015 using multi-gap resistive plate chamber technology, providing a time resolution of 60~ps~\cite{bes3mrpc}.

Simulated data samples produced with a {\sc geant4}-based~\cite{geant} Monte Carlo (MC) package, which includes the geometric description of the BESIII detector and the detector response, are used to determine detection efficiencies and to estimate backgrounds. The simulation models the beam energy spread and initial-state radiation (ISR) in the $e^+e^-$ annihilations with the {\sc kkmc} generator~\cite{kkmc}. The inclusive MC sample includes the production of open charm processes, the ISR production of vector charmonium(-like) states, and the continuum processes incorporated in {\sc kkmc}~\cite{kkmc}. All particle decays are modeled with {\sc evtgen}~\cite{evtgen} using branching fractions either taken from the Particle Data Group (PDG)~\cite{pdg2022}, when available, or otherwise estimated with {\sc lundcharm}~\cite{lundcharm}. Final-state radiation from charged final state particles is incorporated using the {\sc photos} package~\cite{photos}. The input cross section line shape of	$e^+e^-\to D^{*\pm}_sD^{\mp}_s$ is based on the cross section measurement in the energy range from threshold to 4.7 GeV.

\section{BRANCHING FRACTION MEASUREMENT}
At $E_{\rm cm}$ between 4.128~GeV and 4.226 GeV, the $D_s$ mesons are produced dominantly by the $e^+e^-\to D^{*\pm}_sD^\mp_s$ reaction. Therefore, the double-tag (DT) technique ~\cite{markiii} is employed in our selection of $D_s^+\to \tau^+\nu_\tau$ via $\tau^+\to\pi^+\bar{\nu}_\tau$ decays. In this method, single-tag (ST) event is defined in which a $D_s^-$ meson is fully reconstructed via any of thirteen hadronic decay modes and a further DT event is selected by reconstructing the transition $\gamma(\pi^0)$ from the $D_s^{*}$ decay and the $\pi^+$ from the $D_s^+\to \tau^+\nu_\tau$ decay via $\tau^+\to\pi^+\bar{\nu}_\tau$. The $D_s^-$ reconstructed in the ST event can be directly from the reaction $e^+e^-\to D^{*+}_sD_s^-$ (direct tag) or indirectly from the decay of the $D_s^{*-}$ in the conjugate mode (indirect tag). Both direct and indirect tag events are used in further analysis.

For a specific tag mode $i$, the ST yield is given by
\begin{equation}
  N^i_\text{ST} = 2N_{D^{*\pm}_sD^\mp_s}\b^i_\text{ST}\epsilon^i_\text{ST},
\end{equation}
\noindent where $N_{D^{*\pm}_sD^\mp_s}$ is the number of the $D^{*\pm}_sD^\mp_s$ meson pairs produced in the data sample, $\b^i_\text{ST}$ is the branching fraction of the tag mode, and $\epsilon^i_\text{ST}$ is the efficiency for reconstruction of this mode. The factor of 2 accounts for the summation of direct and indirect tags.

The DT event is formed adding the transition $\gamma(\pi^0)$ from the $D_s^{*}$ decay and the $\pi^+$ from $\tau^+$ in the $D^+_s\to \tau^+\nu_\tau$ to an ST event. The DT yield is given by

\begin{equation}
  \begin{aligned}
  N_\text{DT}^{\tau\nu,i} & = & 2N_{D^{*\pm}_sD^\mp_s} \b^i_\text{ST} \b(D_s^+\to \tau^+\nu_\tau)\epsilon^{\tau\nu,i}_\text{DT},
  \end{aligned}
\end{equation}
\noindent where the efficiency $\epsilon^{\tau\nu,i}_\text{DT}$ includes $\b(D^{*+}_s\to\gamma(\pi^0) D^+_s)$, but not the branching fraction $\b(\tau^+\to\pi^+\bar{\nu}_\tau)$.

The branching fraction of the signal decay of $D^+_s\to \tau^+\nu_\tau$ is determined by
\begin{equation}\label{eq:dteqtau}
  \small
  \b(D_s^+\to\tau^+\nu_\tau) = \frac{N_\text{DT}^{\tau\nu}}{\sum_{i}{N^i_\text{ST}(\epsilon^{\tau\nu,i}_\text{DT}/\epsilon^i_\text{ST})} \b(\tau^+\to\pi^+\bar{\nu}_\tau)},
\end{equation}

The systematic uncertainties associated with the ST analysis are largely canceled out in the ratios of $N_\text{DT}^{\tau\nu}/N^i_\text{ST}$ and $\epsilon^{\tau\nu,i}_\text{DT}/\epsilon^i_\text{ST}$. However, there may be a residual uncertainty arising from potential variations in ST reconstruction efficiencies, called tag bias, as discussed in Sec.~\ref{sec:systeff}.

\subsection{\boldmath Selection of $D_s^-$ candidates in ST events}\label{sec:STselect}
Thirteen hadronic decay channels shown in Table~\ref{tab:xyzstyields} are used as the tag modes, where the intermediate particles are reconstructed as
$\pi^0\to\gamma\gamma$,
$K^0_S\to\pi^+\pi^-$,
$\eta\to\gamma\gamma$,
$\eta_{3\pi}\to\pi^+\pi^-\pi^0$,
$\rho^{-(0)}\to\pi^{0(+)}\pi^-$,
$\eta^\prime_{\pi\pi\eta}\to\pi^+\pi^-\eta$, and
$\eta^\prime_{\gamma\rho}\to\gamma\rho^0$. These modes are selected after performing the full analysis procedure on simulated data samples, with the aim of maximizing the signal sensitivity while introducing minimum bias on the measurement.

The selection criteria for $D_s^-$ daughters and the reconstruction procedures are the same as those described in Refs.~\cite{zhangsf,BES3taupipi0}. Tracks must be within the fiducial region ($|\!\cos\theta| < 0.93$, where $\theta$ is the polar angle defined with respect to the $z$-axis, which is the symmetry axis of the MDC) and originate within $1$~cm ($10$~cm) of the interaction point in the plane transverse to the beam direction (in the beam direction).
This requirement on the primary vertex is not applied for the decays of $K^0_S\to\pi^+\pi^-$, for which the distances of the closest approach of the two charged pions to the interaction point are required to be less than 20~cm along the MDC axis. In addition, the charged-pion pair is constrained to have a common vertex with a loose fit-quality requirement of $\chi^2 < 200$ and the invariant mass of the $\pi^+\pi^-$ combination is required to be within $(0.486,\,0.510)$\,GeV$/c^{2}$.

The $K/\pi$ particle identification (PID) is performed by using the TOF and ${\rm d}E/{\rm d}x$ information. Each charged track is assigned as a pion or kaon if the corresponding hypothesis has a higher likelihood. No PID is performed on the charged pions from the intermediate decay $K^0_S\to\pi^+\pi^-$. In addition, the reconstructed momentum for any charged or neutral pion is required to be greater than $0.1$~GeV$/c$ to suppress events from $D^*\to D\pi$ decays.

Photon candidates are chosen from EMC showers unassociated with any charged track~\cite{2009MAblikimDet}. The shower must start between $0$ and $700$~ns after a beam crossing to suppress electronic noise and showers unrelated to the event. When forming $\pi^0$ and $\eta$ candidates, the showers must have an energy greater than $25$~MeV if they are detected in the barrel EMC and $50$~MeV for the end-cap EMC. The $\pi^0$ and $\eta$ candidates are formed by photon pairs with invariant masses lying within the intervals $(0.115,\,0.150)$\,GeV$/c^{2}$ and $(0.500,\,0.570)$\,GeV$/c^{2}$, respectively. To improve momentum resolution and suppress background, a kinematic fit is imposed on each pair of selected photons to constrain its invariant mass to the known $\pi^{0}$ or $\eta$ mass~\cite{pdg2022}. The $\chi^2$ of this kinematic fit is required to be less than 20.

For the tag modes $D_s^-\to \pi^-\eta$ and $D_s^-\to \rho^-\eta$, the $\eta$ candidates are also formed with the $\pi^+\pi^-\pi^0$ combinations with invariant masses within the interval $(0.530,\,0.570)~\mathrm{GeV}/c^2$. The $\eta^\prime$ candidates are formed from $\pi^+\pi^-\eta$ and $\gamma\rho^0$ combinations with invariant masses lying within the intervals $(0.946,\,0.970)~\mathrm{GeV}/c^2$ and $(0.940,\,0.976)~\mathrm{GeV}/c^2$, respectively. In addition, the minimum energy of the $\gamma$ from $\eta^\prime\to\gamma\rho^0$ decay must be greater than 0.1\,GeV. The $\rho^0$ and $\rho^+$ candidates are reconstructed from $\pi^+\pi^-$ and $\pi^+\pi^0$ combinations with invariant masses within the interval $(0.570,\,0.970)~\mathrm{GeV}/c^2$.

\begin{figure}[htbp]\centering
  \includegraphics[keepaspectratio=true,width=3.4in,angle=0]{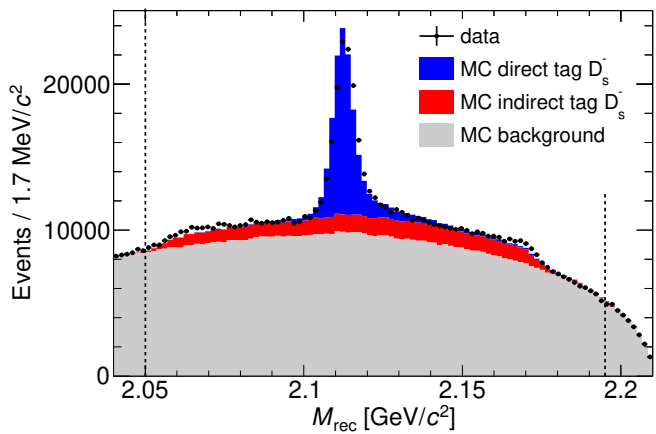}
  \caption{The $M_{\rm rec}$ distribution of the candidates for $D_s^-\to K^-K^+\pi^-$ in the data collected at $E_{\rm cm} = 4.178$~GeV and MC-simulated background. The dashed vertical lines show the ST signal region. The black points with error bars are data, and the solid-filled histograms show the direct tag events (blue), the indirect tag events (red) and simulated backgrounds derived from the inclusive MC sample~(gray).}
  \label{fig:strecoil4180}
\end{figure}

Once the ST event is reconstructed, the recoil mass against the $D_s^-$ tag is calculated as $M_{\text{rec}}^2c^4 = \Big (E_{\text{cm}} - \sqrt{|\vec{p}_{D_s^-}|^2c^2+m^2_{D_s^-}c^4}\Big )^2 - |\vec{p}_{D_s^-}|^2c^2$ in the center-of-mass system of the initial $e^+e^-$, where $\vec{p}_{D_s^-}$ is the three-momentum of the reconstructed $D_s^-$, and $m_{D_s^-}$ is the nominal $D_s^-$ mass~\cite{pdg2022}. Figure~\ref{fig:strecoil4180} shows the $M_{\rm rec}$ distribution for $D_s^-\to K^-K^+\pi^-$ tagged decays in the data collected at $E_{\rm cm} = 4.178$~GeV. All $e^+e^-\to D^{*\pm}_sD^\mp_s$ events accumulate near $m_{D_s^*} = 2.1122$~GeV$/c^2$~\cite{pdg2022}, with the direct tag events populating the central peak and the indirect tag events distributed more broadly. The fraction of both direct and indirect tag events is approximately half.

To select $e^+e^-\to D^{*\pm}_sD^\mp_s$, the $M_{\rm rec}$ of the tagged $D_s^-$ in ST events is required to satisfy the $E_{\text{cm}}$-dependent requirement listed in Table~\ref{tab:mreccut}. This requirement retains most of the $D_s^-$ mesons from $e^+e^-\to D^{*\pm}_sD^\mp_s$, and maintain roughly constant tag efficiencies for different $E_{\text{cm}}$ datasets. The tails of the indirect tag events exhibit a wider extension. A tight requirement is applied to the data collected at $E_{\rm cm} = 4.226$~GeV, as its energy surpasses the threshold for the production of $D_s^{*}D_s^{*}$. When multiple reconstructed candidates are found for a given $D_s^-$ tag mode and electric charge, only the one with the $M_{\rm rec}$ closest to the nominal $D_s^{*+}$ mass~\cite{pdg2022} is kept for further analysis.

\begin{figure*}
\begin{tikzpicture}
\node [ above right, inner sep=0] (image) at (0,0) {\includegraphics[keepaspectratio=true,width=\textwidth,angle=0]{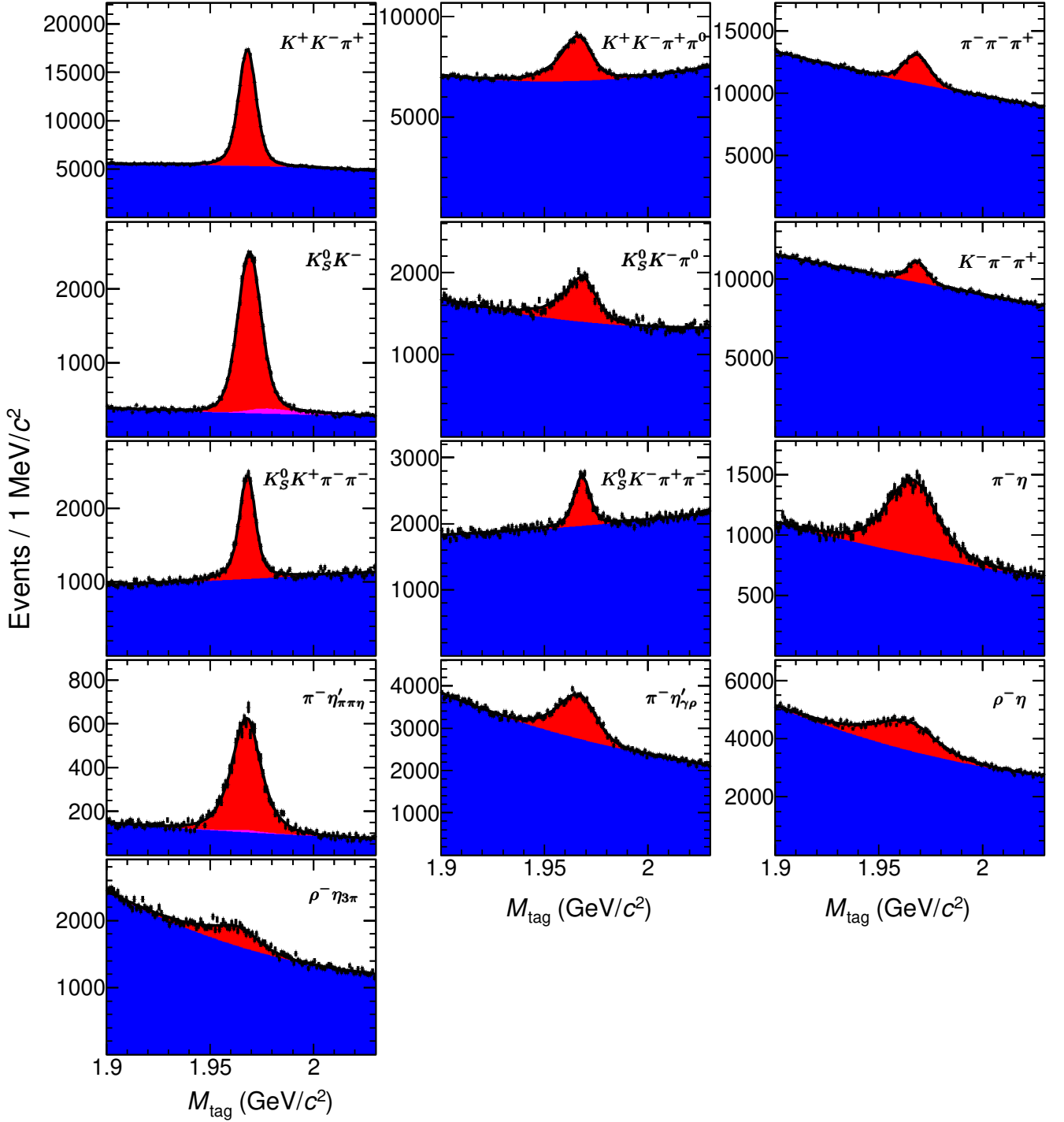}};
\node[align=left] at (5.6,18.8) {$\pmb{K^+K^-\pi^+}$};
\node[align=left] at (11.2,18.8) {$\pmb{K^+K^-\pi^+\pi^0}$};
\node[align=left] at (17,18.8) {$\pmb{\pi^-\pi^-\pi^+}$};
\node[align=left] at (5.8,15.1) {$\pmb{K^0_SK^-}$};
\node[align=left] at (11.3,15.1) {$\pmb{K^0_SK^-\pi^0}$};
\node[align=left] at (17,15.1) {$\pmb{K^-\pi^-\pi^+}$};
\node[align=left] at (5.5,11.4) {$\pmb{K^0_SK^+\pi^-\pi^-}$};
\node[align=left] at (11.2,11.4) {$\pmb{K^0_SK^-\pi^+\pi^-}$};
\node[align=left] at (17.2,11.4) {$\pmb{\pi^-\eta}$};
\node[align=left] at (5.8,7.7) {$\pmb{\pi^-\eta^\prime_{\pi\pi\eta}}$};
\node[align=left] at (11.5,7.7) {$\pmb{\pi^-\eta^\prime_{\gamma\rho}}$};
\node[align=left] at (17.2,7.7) {$\pmb{\rho^-\eta}$};
\node[align=left] at (5.8,4.4) {$\pmb{\rho^-\eta_{3\pi}}$};
\end{tikzpicture}
\caption{Fits to the $M_{\rm tag}$ distributions of the candidates for various ST modes from the data collected at $E_{\rm cm} = 4.178$~GeV. The points with error bars are data, while the black-solid curves represent the total fits, the red solid-filled histograms are the fitted signal shapes and the blue solid-filled histograms are the fitted background shapes. The magenta solid-filled histograms for the $K^+K^-\pi^+$, $K_S^0K^-$ and $\pi^-\eta^\prime_{\pi\pi\eta}$ tag modes are the fitted peaking background shapes of $D^-\to K^+\pi^-\pi^-$, $D^-\to K^0_S\pi^-$ and $D_s^-\to \pi^-\pi^+\pi^-\eta$. For each fit, the $\chi^2$/NDOF ranges from 0.8 to 1.3, where NDOF is the number of degrees of freedom.}
\label{fig:STFIT_DATA}
\end{figure*}

Figure~\ref{fig:STFIT_DATA} shows the $M_{\rm tag}$ distributions of the ST events selected from the data collected at $E_{\rm cm} = 4.178$~GeV as example. The ST yield for each tag mode is determined from an unbinned maximum likelihood fit to the corresponding $M_{\rm tag}$ distribution in the range of $1.90<M_{\rm tag}<2.03$~GeV$/c^2$, where $M_{\rm tag}$ is the mass of the reconstructed $D_s^-$ candidate in ST event. The signal shapes are derived from MC simulation, obtained by the Gaussian kernel estimation method~\cite{keypdf}, and convolved with a Gaussian function to account for the resolution difference between data and MC simulation. For the tag modes $D_s^-\to K^-K^+\pi^-$, $D_s^-\to K_S^0K^-$ and $D_s^-\to \pi^-\eta^\prime_{\pi\pi\eta}$, the peaking backgrounds from $D^-\to K^+\pi^-\pi^-$, $D^-\to K^0_S\pi^-$ and $D_s^-\to \pi^-\pi^+\pi^-\eta$ are described by individual simulated shapes convolved with the same Gaussian function used in the signal shape, and the sizes of the $D_s^-\to K^-K^+\pi^-$ and $D_s^-\to K_S^0K^-$ are free while the size of the $D_s^-\to \pi^-\eta^\prime_{\pi\pi\eta}$ is fixed based on MC simulation. The non-peaking background is modeled using a Chebyshev polynomial function of order one to three, which has been validated by analyzing the inclusive sample.

The fit results are also shown on Fig.~\ref{fig:STFIT_DATA} for the data collected at $E_{\rm cm} = 4.178$~GeV.
The obtained $\text{ST}$ yields from data for various tag modes and data samples within the $M_{\rm tag}$ window are shown in Table~\ref{tab:xyzstyields}. Additionally, in Table~\ref{tab:xyzSTeff}, the ST efficiencies for various tag modes are illustrated, obtained through the analysis of the inclusive MC sample within the $M_{\rm tag}$ window.

\begin{table}[htb]\centering
  \caption{The requirement on $M_{\rm rec}$ for the ST candidates at each datasets collected at different $E_{\rm cm}$.}
  \label{tab:mreccut}
    \begin{tabular}{C{2cm} C{3.5cm}}
    \hline\hline
    $E_{\rm cm}$ (GeV) & $M_{\rm rec}$~(GeV$/c^2$) \\\hline
    4.128    & $(2.060,\,2.155)$ \\
    4.157    & $(2.054,\,2.175)$ \\
    4.178    & $(2.050,\,2.195)$ \\
    4.189    & $(2.048,\,2.205)$ \\
    4.199    & $(2.046,\,2.215)$ \\
    4.209    & $(2.044,\,2.225)$ \\
    4.219    & $(2.042,\,2.235)$ \\
    4.226    & $(2.040,\,2.220)$ \\
    \hline\hline
    \end{tabular}
\end{table}

\begin{table*}[htb]\centering
  \caption{The $\text{ST}$ yields ($N_\text{ST}$) for each tag mode and data sample, in units of $10^3$. ``SUM'' denotes the total ST yield summed over tag modes. The uncertainties are statistical only.}
  \label{tab:xyzstyields}
  \scalebox{1.0}
  {
    \begin{tabular}{c| c c c c c c c c}
    \hline\hline
    \diagbox{Tag mode}{$E_\text{cm}$ (GeV)} & $4.128$ & $4.157$ & $4.178$ & $4.189$ & $4.199$ & $4.209$ &  $4.219$ &  $4.226$ \\
    \hline
    {\ensurestackMath{
    \alignCenterstack{
    K^{+}K^{-}\pi^{-}\cr
    K^{+}K^{-}\pi^{-}\pi^{0}\cr
    \pi^{+}\pi^{-}\pi^{-}\cr
    K_{S}^{0}K^{-}\cr
    K_{S}^{0}K^{-}\pi^{0}\cr
    K^{-}\pi^{+}\pi^{-}\cr
    K_{S}^{0}K^{+}\pi^{-}\pi^{-}\cr
    K_{S}^{0}K^{-}\pi^{+}\pi^{-}\cr
    \pi^{-}\eta\cr
    \pi^{-}\eta^{\prime}_{\pi\pi\eta}\cr
    \pi^{-}\eta^{\prime}_{\gamma\rho}\cr
    \rho^{-}\eta\cr
    \rho^{-}\eta_{3\pi}}}} &
    {\ensurestackMath{
    \alignCenterstack{
    11.6\pm0.2\cr
    4.0\pm0.4\cr
    2.5\pm0.2\cr
    2.5\pm0.1\cr
    0.9\pm0.1\cr
    1.8\pm0.3\cr
    1.3\pm0.1\cr
    0.8\pm0.1\cr
    1.6\pm0.1\cr
    0.8\pm0.0\cr
    2.0\pm0.2\cr
    3.4\pm0.4\cr
    1.0\pm0.2}}} &
    {\ensurestackMath{
    \alignCenterstack{
    18.2\pm0.2\cr
    6.4\pm0.4\cr
    5.0\pm0.4\cr
    4.2\pm0.1\cr
    1.5\pm0.1\cr
    2.2\pm0.2\cr
    1.9\pm0.1\cr
    1.0\pm0.1\cr
    2.7\pm0.2\cr
    1.3\pm0.1\cr
    2.9\pm0.3\cr
    5.8\pm0.6\cr
    1.4\pm0.3}}} &
    {\ensurestackMath{
    \alignCenterstack{
    146.2\pm0.7\cr
    49.8\pm1.1\cr
    41.1\pm1.0\cr
    32.2\pm0.3\cr
    12.8\pm0.4\cr
    18.5\pm0.7\cr
    16.8\pm0.3\cr
    9.0\pm0.3\cr
    20.6\pm1.0\cr
    10.2\pm0.0\cr
    25.8\pm0.8\cr
    42.0\pm2.2\cr
    10.6\pm1.0}}} &
    {\ensurestackMath{
    \alignCenterstack{
    25.1\pm0.3\cr
    8.8\pm0.5\cr
    7.1\pm0.5\cr
    5.6\pm0.1\cr
    2.2\pm0.2\cr
    3.6\pm0.4\cr
    2.9\pm0.1\cr
    1.5\pm0.1\cr
    3.2\pm0.2\cr
    1.8\pm0.1\cr
    4.0\pm0.2\cr
    6.0\pm0.7\cr
    1.8\pm0.3}}} &
    {\ensurestackMath{
    \alignCenterstack{
    22.9\pm0.3\cr
    8.0\pm0.5\cr
    6.3\pm0.5\cr
    5.1\pm0.1\cr
    2.6\pm0.0\cr
    2.9\pm0.1\cr
    2.7\pm0.1\cr
    1.6\pm0.2\cr
    3.2\pm0.2\cr
    1.5\pm0.1\cr
    3.6\pm0.3\cr
    7.1\pm0.5\cr
    1.7\pm0.6}}} &
    {\ensurestackMath{
    \alignCenterstack{
    23.9\pm0.3\cr
    7.7\pm0.5\cr
    6.4\pm0.5\cr
    5.0\pm0.1\cr
    2.0\pm0.0\cr
    2.6\pm0.2\cr
    2.4\pm0.1\cr
    1.4\pm0.2\cr
    3.4\pm0.2\cr
    1.6\pm0.1\cr
    4.3\pm0.4\cr
    6.6\pm0.7\cr
    2.4\pm0.6}}} &
    {\ensurestackMath{
    \alignCenterstack{
    20.8\pm0.3\cr
    6.9\pm0.5\cr
    4.9\pm0.5\cr
    4.3\pm0.1\cr
    1.5\pm0.2\cr
    2.4\pm0.3\cr
    2.1\pm0.1\cr
    1.3\pm0.2\cr
    2.6\pm0.2\cr
    1.4\pm0.1\cr
    3.9\pm0.4\cr
    5.0\pm0.5\cr
    2.5\pm1.4}}} &
    {\ensurestackMath{
    \alignCenterstack{
    31.4\pm0.4\cr
    10.7\pm0.6\cr
    8.0\pm1.2\cr
    6.9\pm0.1\cr
    2.7\pm0.2\cr
    5.2\pm0.5\cr
    3.2\pm0.1\cr
    1.6\pm0.2\cr
    4.5\pm0.5\cr
    2.2\pm0.1\cr
    5.0\pm0.4\cr
    9.7\pm1.3\cr
    2.8\pm0.3}}} \\
    \hline
    SUM &
    {\ensurestackMath{\alignCenterstack{~34.2\pm0.8}}} &
    {\ensurestackMath{\alignCenterstack{~54.6\pm1.0}}} &
    {\ensurestackMath{\alignCenterstack{~435.8\pm3.3}}} &
    {\ensurestackMath{\alignCenterstack{~73.7\pm1.2}}} &
    {\ensurestackMath{\alignCenterstack{~69.1\pm1.2}}} &
    {\ensurestackMath{\alignCenterstack{~69.8\pm1.3}}} &
    {\ensurestackMath{\alignCenterstack{~59.5\pm1.8}}} &
    {\ensurestackMath{\alignCenterstack{~94.1\pm2.2}}} \\
    \hline\hline
    \end{tabular}
  }
\end{table*}

\begin{table*}[htb]\centering
  \caption{The $\text{ST}$ efficiencies ($\epsilon_\text{ST}$ in $\%$) for each tag mode and data sample. The uncertainties are statistical only. Efficiencies do not include the branching fractions of the intermediate decays $K^0_S\to\pi^+\pi^-$, $\pi^0\to\gamma\gamma$, $\eta\to\gamma\gamma$, $\eta_{3\pi}\to\pi^+\pi^-\pi^0$, $\eta^\prime_{\pi\pi\eta}\to\pi^+\pi^-\eta$, $\eta^\prime_{\gamma\rho}\to\gamma\rho^0$, or $\rho\to\pi\pi$.
  }
  \label{tab:xyzSTeff}
  \scalebox{1.00}
  {
    \begin{tabular}{c| c c c c c c c c}
    \hline\hline
    \diagbox{Tag mode}{$E_\text{cm}$ (GeV)} & $4.128$ & $4.157$ & $4.178$ & $4.189$ & $4.199$ & $4.209$ &  $4.219$ &  $4.226$ \\
    \hline
    {\ensurestackMath{
    \alignCenterstack{
    K^{+}K^{-}\pi^{-}\cr
    K^{+}K^{-}\pi^{-}\pi^{0}\cr
    \pi^{+}\pi^{-}\pi^{-}\cr
    K_{S}^{0}K^{-}\cr
    K_{S}^{0}K^{-}\pi^{0}\cr
    K^{-}\pi^{+}\pi^{-}\cr
    K_{S}^{0}K^{+}\pi^{-}\pi^{-}\cr
    K_{S}^{0}K^{-}\pi^{+}\pi^{-}\cr
    \pi^{-}\eta\cr
    \pi^{-}\eta^{\prime}_{\pi\pi\eta}\cr
    \pi^{-}\eta^{\prime}_{\gamma\rho}\cr
    \rho^{-}\eta\cr
    \rho^{-}\eta_{3\pi}}}} &
    {\ensurestackMath{
    \alignCenterstack{
    44.9\pm0.1\cr
    13.6\pm0.2\cr
    59.5\pm0.6\cr
    49.6\pm0.2\cr
    19.3\pm0.4\cr
    51.6\pm1.1\cr
    22.7\pm0.2\cr
    20.3\pm0.4\cr
    51.6\pm0.6\cr
    26.1\pm0.5\cr
    34.5\pm0.5\cr
    21.7\pm0.3\cr
    9.8\pm0.3}}} &
    {\ensurestackMath{
    \alignCenterstack{
    44.6\pm0.1\cr
    13.7\pm0.1\cr
    60.1\pm0.5\cr
    49.6\pm0.2\cr
    19.0\pm0.3\cr
    51.0\pm0.8\cr
    23.0\pm0.2\cr
    20.3\pm0.3\cr
    51.6\pm0.4\cr
    25.6\pm0.3\cr
    34.5\pm0.3\cr
    21.2\pm0.2\cr
    9.9\pm0.2}}} &
    {\ensurestackMath{
    \alignCenterstack{
    43.7\pm0.1\cr
    13.9\pm0.1\cr
    57.8\pm0.2\cr
    49.6\pm0.1\cr
    19.5\pm0.1\cr
    50.6\pm0.3\cr
    23.6\pm0.1\cr
    21.4\pm0.1\cr
    51.7\pm0.2\cr
    25.5\pm0.1\cr
    33.8\pm0.1\cr
    20.9\pm0.1\cr
    9.9\pm0.1}}} &
    {\ensurestackMath{
    \alignCenterstack{
    43.8\pm0.1\cr
    14.0\pm0.1\cr
    56.6\pm0.4\cr
    49.5\pm0.2\cr
    19.4\pm0.3\cr
    51.9\pm0.7\cr
    23.7\pm0.1\cr
    21.1\pm0.3\cr
    51.1\pm0.4\cr
    25.1\pm0.3\cr
    33.5\pm0.3\cr
    21.4\pm0.2\cr
    9.9\pm0.2}}} &
    {\ensurestackMath{
    \alignCenterstack{
    43.9\pm0.1\cr
    14.1\pm0.1\cr
    57.2\pm0.4\cr
    49.4\pm0.2\cr
    20.0\pm0.3\cr
    50.9\pm0.7\cr
    23.8\pm0.2\cr
    21.4\pm0.3\cr
    51.4\pm0.4\cr
    24.8\pm0.3\cr
    34.3\pm0.3\cr
    20.9\pm0.2\cr
    9.5\pm0.2}}} &
    {\ensurestackMath{
    \alignCenterstack{
    43.6\pm0.1\cr
    14.1\pm0.1\cr
    56.1\pm0.4\cr
    48.9\pm0.2\cr
    19.1\pm0.3\cr
    50.1\pm0.7\cr
    23.5\pm0.1\cr
    21.0\pm0.3\cr
    51.1\pm0.4\cr
    25.2\pm0.3\cr
    33.9\pm0.3\cr
    20.8\pm0.2\cr
    9.4\pm0.2}}} &
    {\ensurestackMath{
    \alignCenterstack{
    43.2\pm0.1\cr
    14.1\pm0.1\cr
    55.5\pm0.5\cr
    48.9\pm0.2\cr
    19.7\pm0.3\cr
    48.1\pm0.9\cr
    23.5\pm0.2\cr
    21.0\pm0.3\cr
    50.4\pm0.5\cr
    25.1\pm0.4\cr
    32.8\pm0.4\cr
    20.5\pm0.3\cr
    9.4\pm0.3}}} &
    {\ensurestackMath{
    \alignCenterstack{
    43.6\pm0.1\cr
    14.2\pm0.1\cr
    57.3\pm0.5\cr
    49.4\pm0.1\cr
    19.8\pm0.3\cr
    50.3\pm0.7\cr
    23.9\pm0.1\cr
    21.4\pm0.3\cr
    51.0\pm0.4\cr
    25.5\pm0.3\cr
    34.5\pm0.3\cr
    20.6\pm0.2\cr
    9.9\pm0.2}}} \\
    \hline\hline
  \end{tabular}
  }
\end{table*}

\subsection{\boldmath Selection of transition $\gamma(\pi^0)$ from $D^*_s$}\label{sec:softgamma}
In the presence of tagged $D_s^-$, the $D_s^{*+}\to D_s^+ \gamma(\pi^0)$ transition is distinguished from combinatorial backgrounds by a kinematic variable:
\begin{equation}\label{eq:dele}
\Delta E = E_{\rm cm}-E_{\rm tag}-E_{\rm miss}-E_{\gamma(\pi^0)}.
\end{equation}

\noindent Here $E_{\rm miss} = \sqrt{|\vec{p}_{\rm miss}|^2c^2+m_{D_s^+}^2c^4}$ and $\vec{p}_{\rm miss} = -\vec{p}_{D_s^-}-\vec{p}_{\gamma(\pi^0)}$ are the missing energy and momentum of the recoiling system of the transition $\gamma(\pi^0)$ and the tagged $D_s^-$, respectively. In case where there are several $\gamma(\pi^0)$ candidate, the one giving the smallest $|\Delta E|$ is kept.

In the $D_s^{*+}$ rest frame, the energy of the transition photon has a monochromatic value of $(m^2_{D_s^*}c^4 - m^2_{D_s}c^4)/(2m_{D_s^*}c^2)=0.1389$~GeV. The four-momenta of the $D_s^*$ candidate are calculated under indirect and direct $D_s^-$ tag hypotheses, which are $\bm{p}_{D^{*+}_s} = \bm{p}_{e^+e^-} - \bm{p}_{D_s^-}$ and $\bm{p}_{D^{*-}_s} = \bm{p}_\gamma + \bm{p}_{D_s^-}$, respectively. Here, $\bm{p}_{e^+e^-}$ is the initial four-momentum of the $e^+e^-$ system. The combination giving closest mass to the nominal $D_s^*$ mass is selected. To further suppress background, the energy of transition photon in the $D_s^{*+}$ rest frame is required to be within $0.114<E_\gamma<0.149$~GeV for both cases. This criterion is optimized by maximizing figure-of-merit (FoM) defined as $S/\sqrt{S+B}$, where $S$ and $B$ are the signal and background event yields from inclusive MC sample. The photon selection efficiency is about 85\%. No similar energy requirement is imposed for $D^*_s\to \pi^0D_s$.

\subsection{\boldmath Selection of $D_s^+\to\tau^+\nu_\tau$}\label{sec:dtana}

The $D_s^+ \rightarrow \tau^+\nu_\tau$ signal candidates are reconstructed via $\tau^+ \to \pi^+\bar{\nu}_\tau$ in events containing the transition $\gamma(\pi^0)$ and the tagged $D_s^-$. We require only one additional track that is not used in the tag reconstruction ($N_{\rm extra}^{\rm char} = 0$) and no additional $\pi^0$ candidates can be formed ($N_{\rm extra}^{\pi^0} = 0$). The particle candidate of the signal must have the opposite sign charge to the tagged $D_s^-$ and satisfy the pion PID criteria described in Sec.~\ref{sec:STselect}.

To suppress background events associated with unreconstructed or misreconstructed particles such as electrons that are misidentified as pions, photons from $\pi^0$ and $\eta$ decays and fake photons misidentified from showers produced by $K^0_L$, the following three additional variables are selected. The first one is the ratio of the energy deposited in the EMC over the MDC momentum of the charged pion from $\tau^+$ decays, labeled as EOP. Because the electron deposits most of its energy in the EMC, the EOP of the $\pi^+$ candidate has to be less than 0.9 to suppress background events with misreconstructed electrons. The second one is the maximum energy of extra photons in events, labeled as $E^\text{max}_\text{neu}$. The $E^\text{max}_\text{neu}$ must be less than 0.3 GeV to ensure that there is no extra energetic photon in the selected events. The third one is $\cos{\theta_\text{miss}}$, where $\theta_\text{miss}$ is the polar angle of $\vec{p}_{\text{miss},\nu} = -\vec{p}_{D_s^-} - \vec{p}_{\gamma(\pi^0)} - \vec{p}_{\pi^+}$ in the $e^+e^-$ center-of-mass frame. To suppress background from $e^{+}e^{-}\to q\bar{q}~(q = u, d, s)$ where no particle is missed, the value of $|\!\cos{\theta_\text{miss}}|$ is required to be less than 0.9 to restrict $\vec{p}_{\text{miss},\nu}$ to point into the fiducial volume of the BESIII detector.

The presence of neutrinos in the final states is inferred from the event missing invariant mass-squared, $M^2_{\rm{miss}} = E^2_{\rm{miss},\nu} - |\vec{p}_{\rm{miss},\nu}c|^2$, where $E_{\rm{miss},\nu} = E_{\rm{cm}} -\sqrt{|\vec{p}_{D_s^-}c|^2+m^2_{D_s^-}c^4} - E_{\gamma(\pi^0)} -E_{\pi^+}$ is calculated in the $e^+e^-$ center-of-mass frame. The selected candidates must satisfy $-0.2<M^2_{\rm{miss}}<0.6$~${\rm GeV}^2/c^4$ in order to suppress background from $e^{+}e^{-}\to q\bar{q}~(q = u, d, s)$ at higher masses as shown in the top left plot of Fig.~\ref{fig:inputvar}.

\subsection{\boldmath Multivariate analysis}\label{sec:furtherselection}

Although the $M^2_{\rm{miss}}$ is the kinematic variable that provides the best discrimination between signal and backgrounds, the search sensitivity can be further improved by incorporating additional kinematic and topological information from the selected events using a multivariate analysis technique known as Boosted Decision Tree (BDT)~\cite{tmva4}.

The BDT is trained to distinguish the signal from the sum of the expected background processes. The selection of BDT input variables is based on maximizing the separation power while avoiding variables that do not significantly improve performance. Starting with $M^2_{\rm{miss}}$ as the initial variable, additional variables are sequentially tested, and the one giving the most significant improvement in separation is kept. This process is repeated until the additional variables do not further improve the performance. The final set includes the variables of $M_{\rm tag}$ and $m^{\rm tag}_{\rm BC}$ from the tag side, $\theta_{\rm miss}^{\gamma}$, $\rm{cos}\theta_{\rm{miss}}$, and $M_{\rm{miss}}^2$ from the neutrino, $E_{\gamma(\pi^0)}$ from the transition $\gamma(\pi^0)$, $E_\gamma^{\rm sum}$ representing the summed energy of extra photons in the event, and $\cos\theta_{\pi^+}$ and $\vec{p}_{\pi^+}$ of the $\pi^+$ from the signal side. Here, $m^{\rm tag}_{\rm BC} = \sqrt{E^2_\text{beam}-|\vec{p}_{D_s^-}|^2c^2}$ is the beam-constrained mass of the $D_s^-$ candidate in ST event, in which $E_\text{beam}$ denotes the beam energy and $\theta_{\rm miss}^{\gamma}$ refers to the opening angle between $\vec{p}_{\rm{miss},\nu}$ and the most energetic photon.

\begin{figure*}[htbp]\centering
  \includegraphics[keepaspectratio=true,width=0.325\textwidth,angle=0]{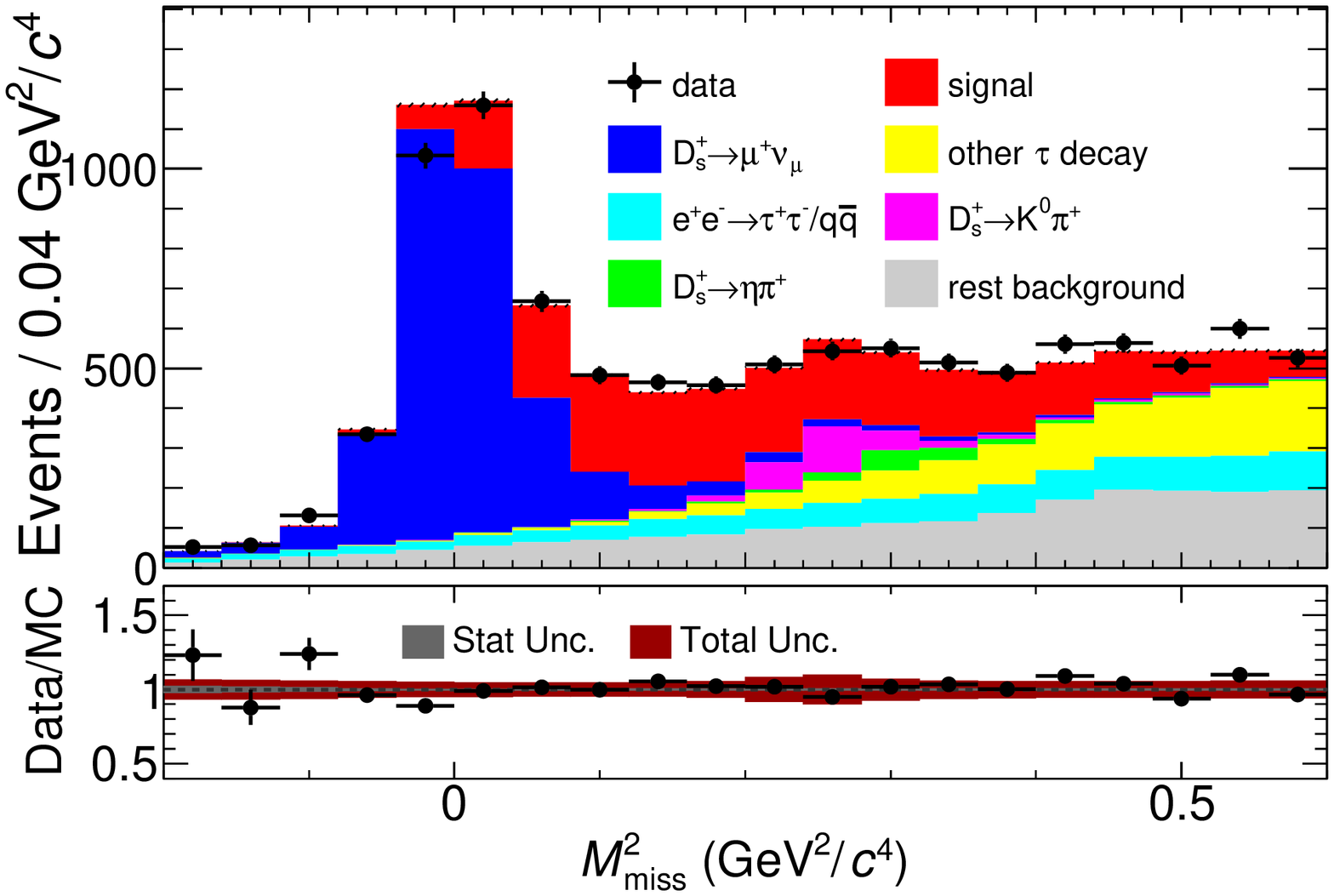}
  \includegraphics[keepaspectratio=true,width=0.325\textwidth,angle=0]{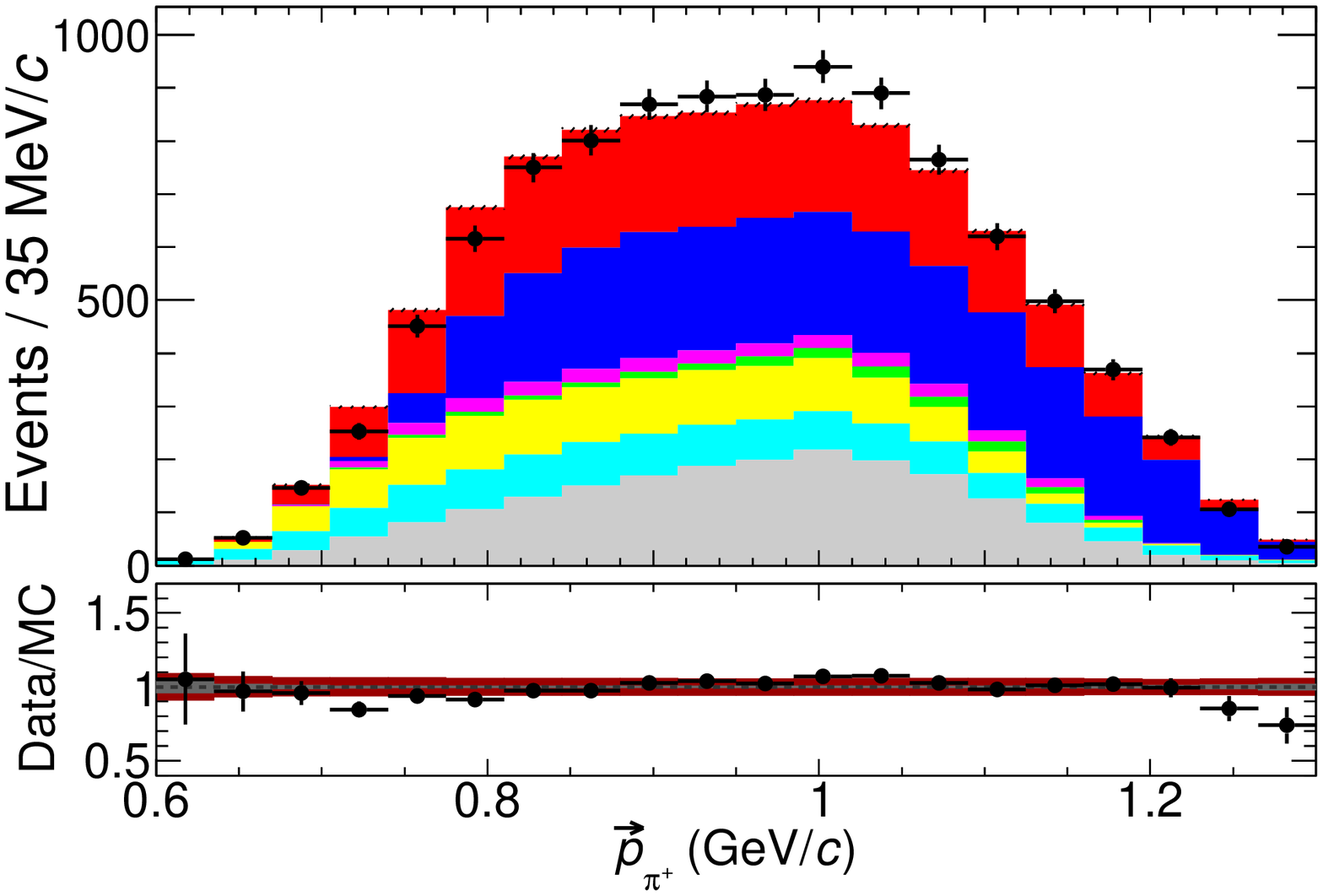}
  \includegraphics[keepaspectratio=true,width=0.325\textwidth,angle=0]{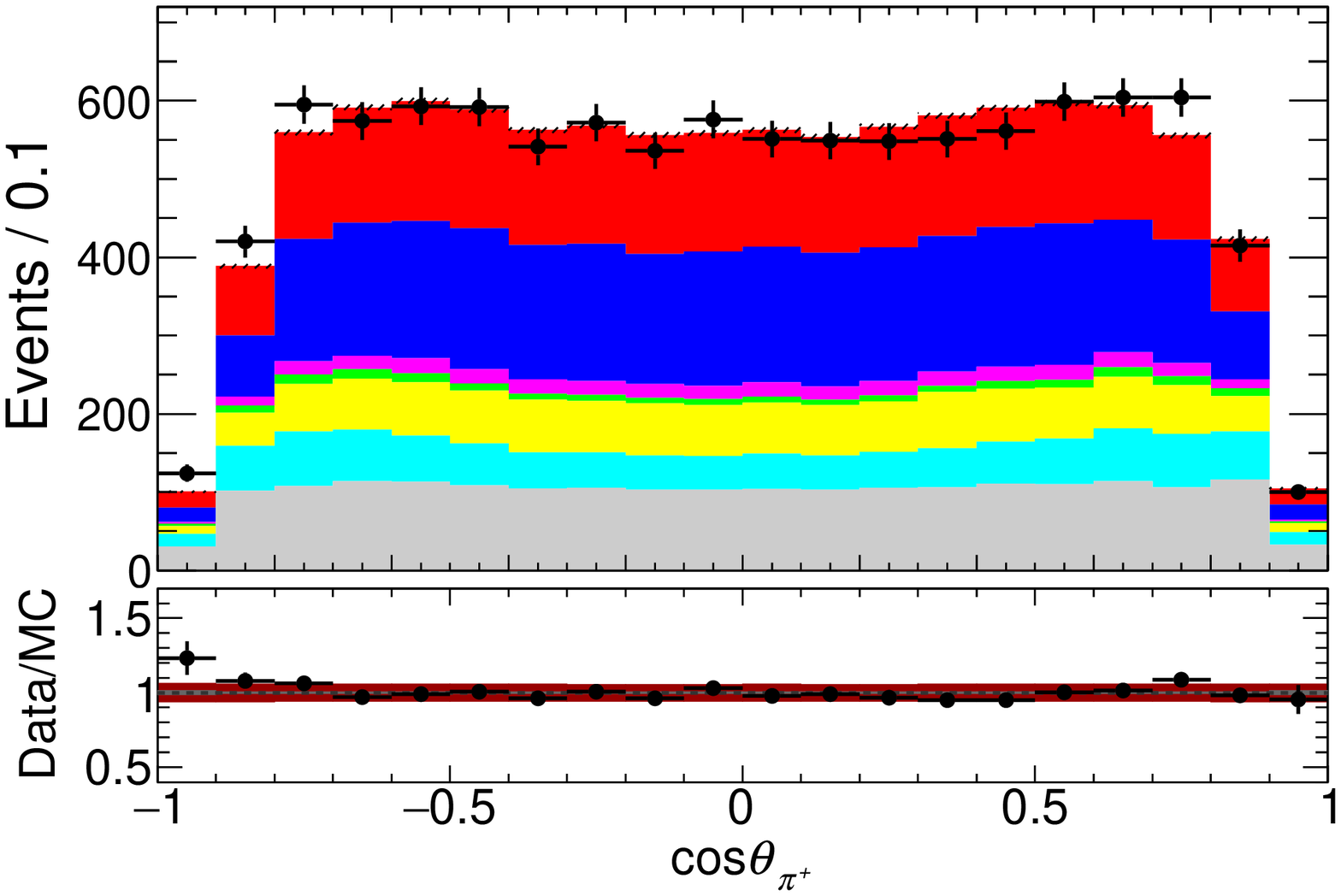}
  \includegraphics[keepaspectratio=true,width=0.325\textwidth,angle=0]{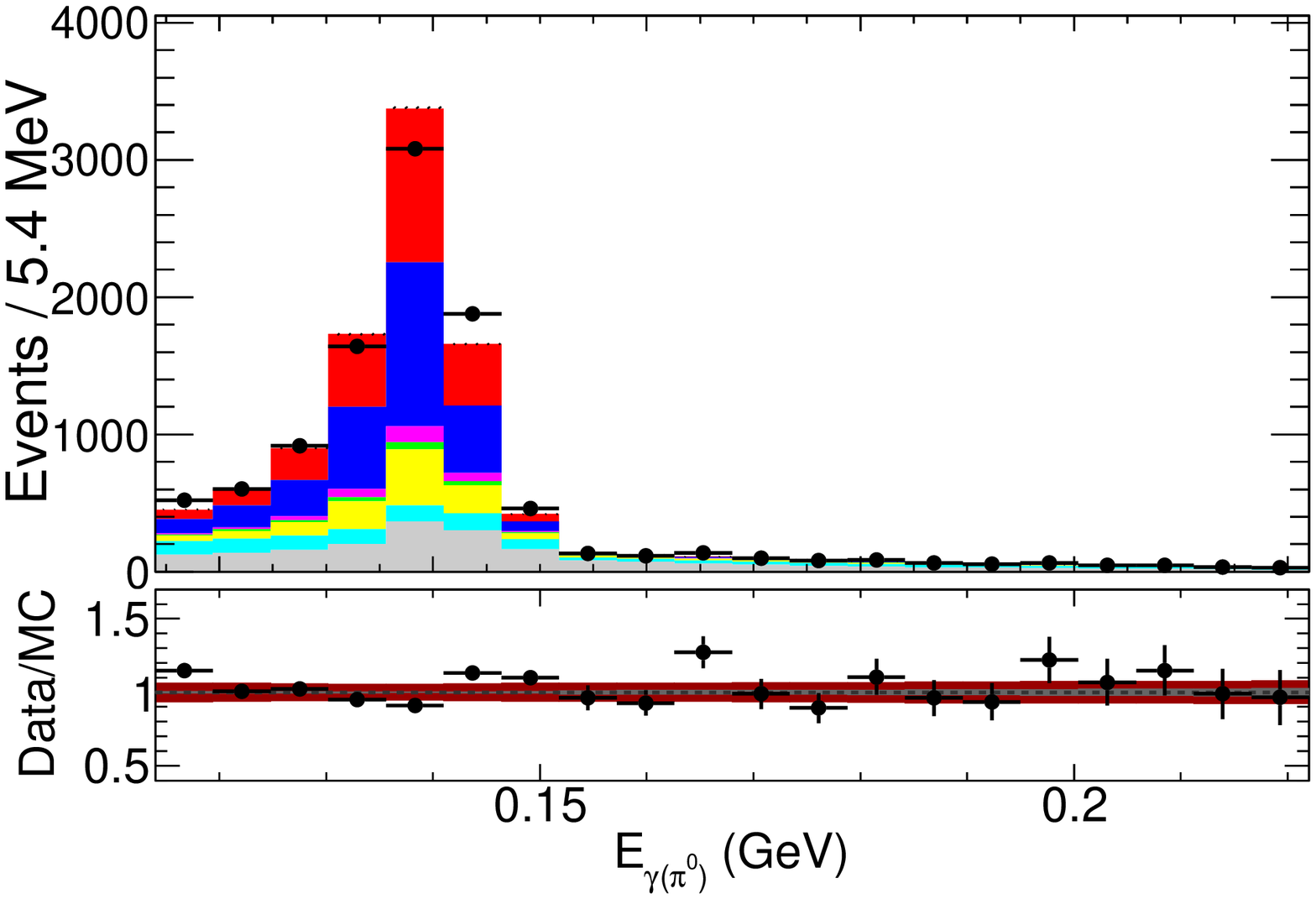}
  \includegraphics[keepaspectratio=true,width=0.325\textwidth,angle=0]{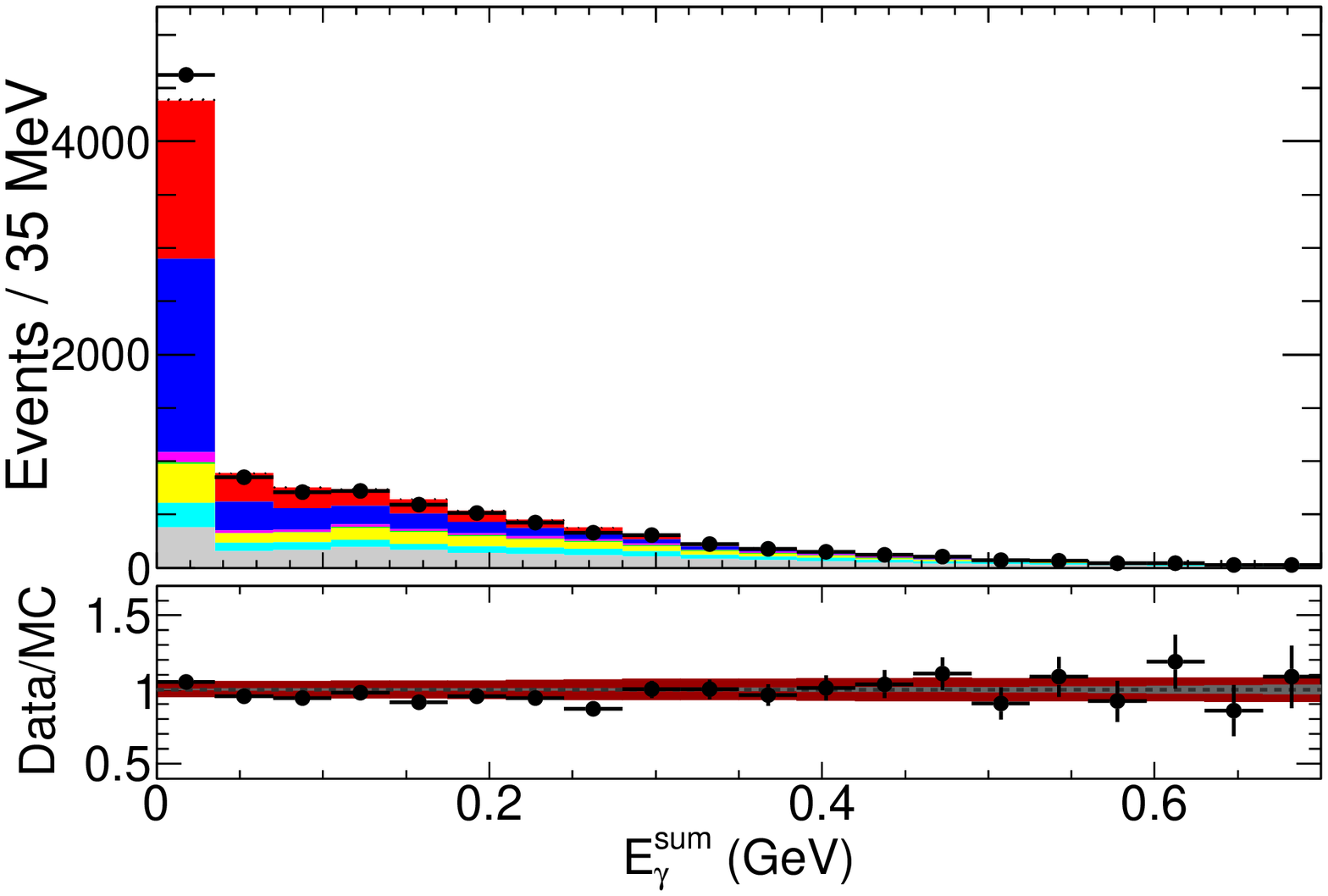}
  \includegraphics[keepaspectratio=true,width=0.325\textwidth,angle=0]{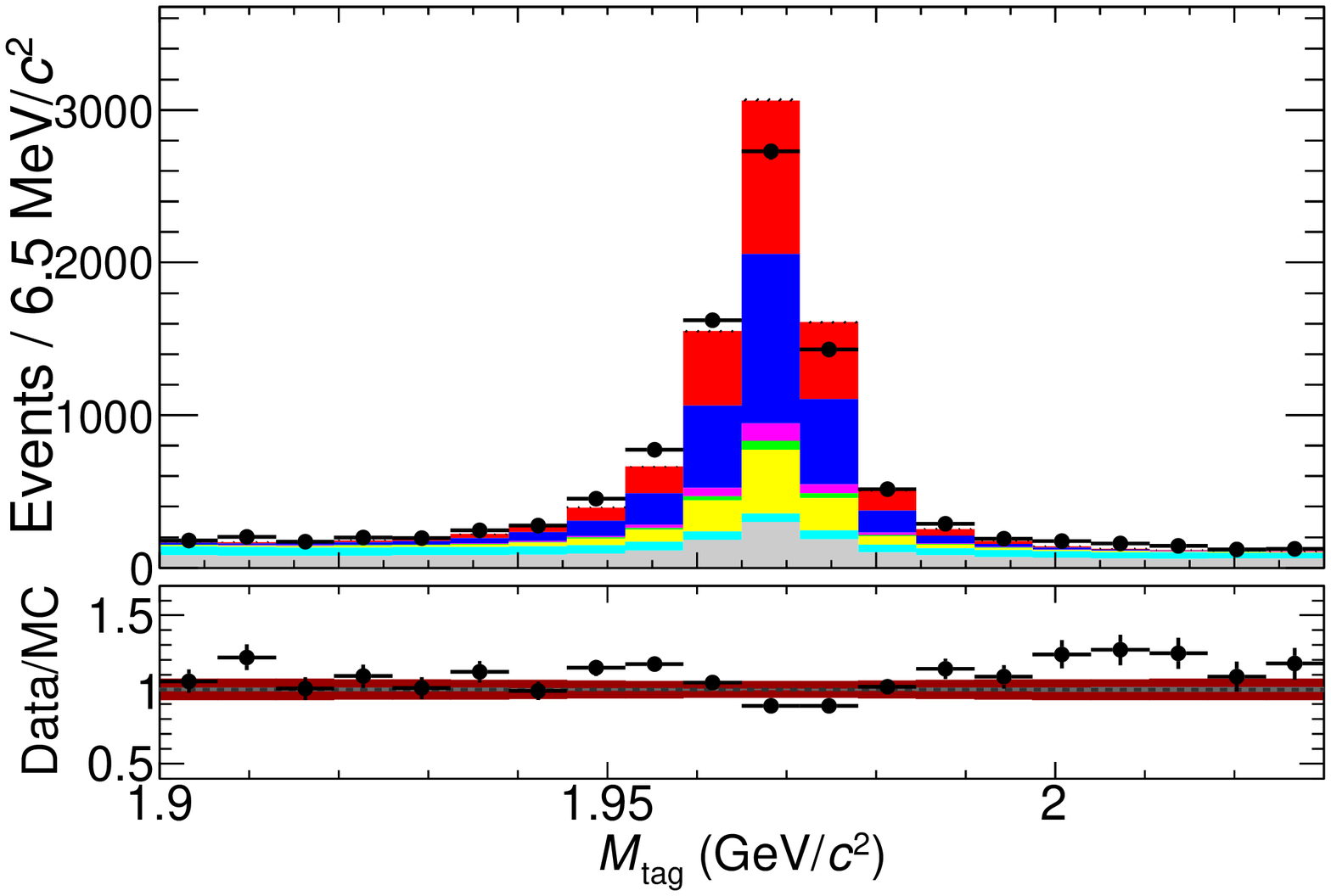}
  \includegraphics[keepaspectratio=true,width=0.325\textwidth,angle=0]{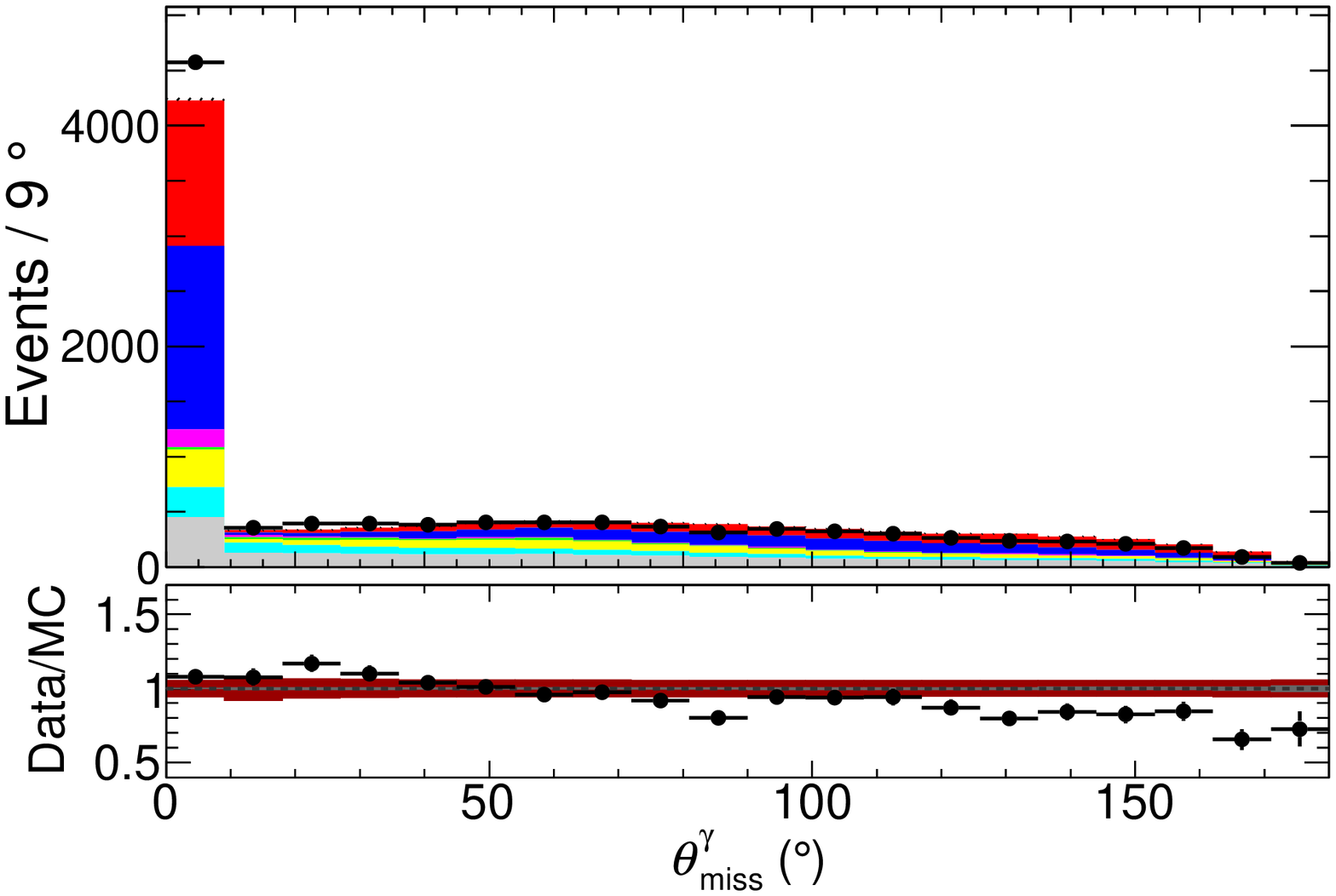}
  \includegraphics[keepaspectratio=true,width=0.325\textwidth,angle=0]{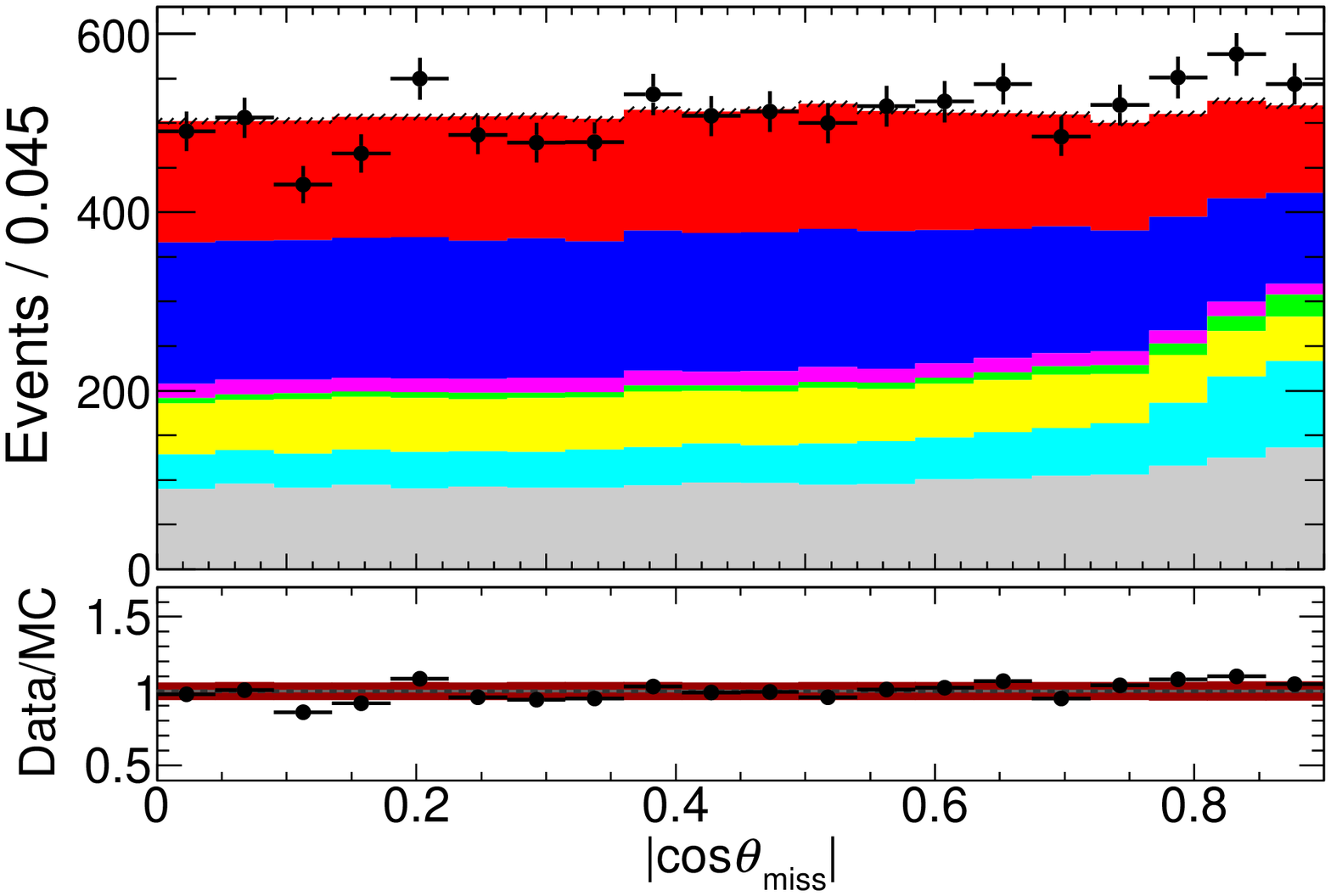}
  \includegraphics[keepaspectratio=true,width=0.325\textwidth,angle=0]{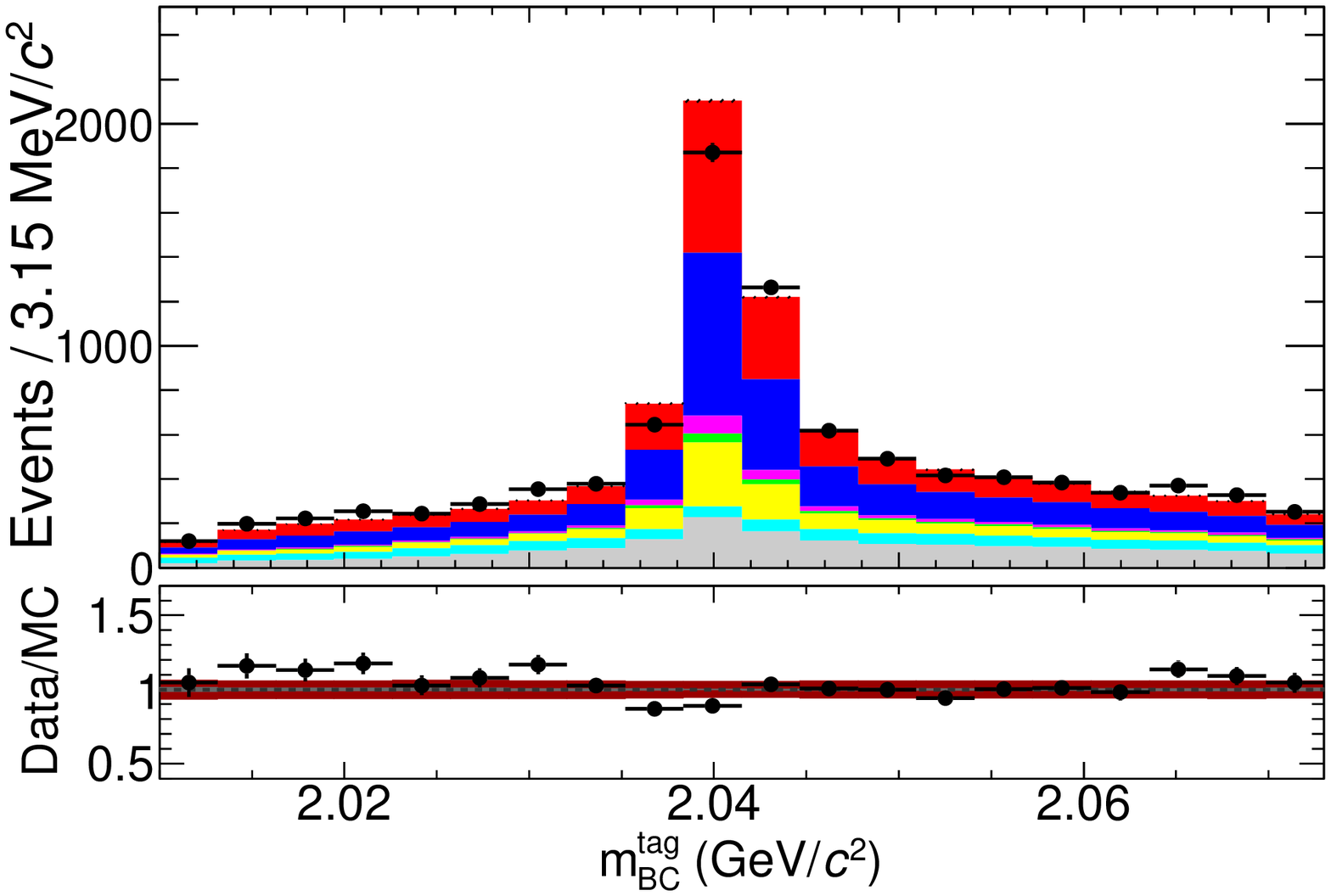}
  \caption{Distributions of various input variables of the BDT. The black points with error bars are data, the red solid-filled histogram shows the signal, the blue solid-filled histogram is the $D_s^+\to \mu^+\nu_{\mu}$ background, the yellow solid-filled histogram is the other $\tau^{+}$ decays background, the cyan solid-filled histogram is the $e^{+}e^{-}\to \tau^{+}\tau^{-}/q \bar{q}~(q = u, d, s)$ background, the magenta solid-filled histogram is the $D_s^+\to K^0\pi^+$ background, the green solid-filled histogram is the $D_s^+\to \eta\pi^+$ background and the gray solid-filled histogram is the remaining background. The legend in the first figure is applicable to all figures.}
  \label{fig:inputvar}
\end{figure*}

Figure~\ref{fig:inputvar} shows the comparisons of the input variables of the BDT between data and MC simulation. It can be seen that the data and MC simulation are in an overall good agreement. Observable data-MC discrepancies in distributions of input variables will be considered as one source of systematic uncertainties as detailed in Sec.~\ref{sec:systfit}. 

The TMVA~\cite{tmva3} framework is used to train the BDT. The values for the hyperparameters are determined by seeking the configuration that offers the best separation between signal and background in a coarsely binned multi-dimensional parameter space defined by the hyperparameters. This is followed by fine-grained one-dimensional scans of individual hyperparameters to ensure an unbiased training and evaluation of the BDT using the complete set of simulated MC events, the MC events are divided into two equal-sized samples, namely A with even event number and B with odd event number. The performance of the BDT trained on sample A (B) is evaluated using sample B (A) to avoid using the same events for both training and evaluation of a particular BDT. The real data is also divided into two parts with even and odd event number, and half of the data is analyzed using the BDT trained on sample A, and the other half using the BDT trained on sample B. Finally, the output distributions of the BDT trained on samples A and B are merged for both the data and simulated events.

\subsection{\boldmath Background composition and modeling}
After the final selection discussed in Sec.~\ref{sec:dtana}, the fractions of remaining background components determined from MC simulations are as follows: $(38.78\pm0.10)\%$ for $D_s^{+}\to\mu^+\nu_\mu$, $(15.31\pm0.06)\%$ for other $\tau$ decays, $(9.33\pm0.05)\%$ for $e^+e^-\to q\bar{q}$, $(3.95\pm0.03)\%$ for $e^+e^-\to\tau^+\tau^-$, $(2.28\pm0.02)\%$ for $D_s^{+}\to\eta\pi^+$, $(2.73\pm0.03)\%$ for $D_s^{+}\to K_L^0K^+$, $(4.16\pm0.03)\%$ for $D_s^{+}\to K^0\pi^+$ and there is approximately 23\% of the background that consists of mixed components, primarily originating from the opencharm processes. Candidates for $D^+_s\to \tau^+\nu_\tau$ with $\tau^+\to \rho^+\bar \nu_\tau$ and $\tau^+\to \mu^+\nu_\mu\bar \nu_\tau$ have been used in the previous BESIII analyses~\cite{BES3taupipi0, BES3taumu} and they will be considered as backgrounds in this measurement. 

Four control regions are defined, orthogonal to the signal region, to validate the modeling of the major backgrounds: 1) $D_s^+\to \mu^+\nu_{\mu}$ as $\mu\nu$ control region, 2) other $\tau^+$ decay as $\tau_{\rm other}$ control region, 3) $e^{+}e^{-}\to\tau^{+}\tau^{-}/q \bar{q}~(q = u, d, s)$ as $qq\tau\tau$ control region and 4) $D_s^+\to \eta\pi^+$ as $\eta\pi$ control region. These control regions are used to check the background modeling, and the selection criterias are summarised in Table~\ref{tab:CR}. Although the proportion of $D^+_s\to\eta \pi^+$ background is small, it forms a peak in $M_{\rm miss}^2$ of our signal region. Nevertheless, a control region can be defined to check its yield and shape. The control region of the $D^+_s\to \eta\pi^+$ background is chosen by using the maximum energy of extra photons. Unlike this process, we do not define the control region of the $D_s^{+}\to K^0\pi^+$ background, since this background is dominated by $D_s^{+}\to K^0_L\pi^+$.


\begin{table*}
  \caption{Definitions of the control regions, where all other selection criteria are imposed except for the corresponding requirements to be shown. The $M_{\rm tag}$ signal and sideband regions are defined to be within and outside $\pm 3\sigma$ around the nominal $D_s$ mass, respectively.  See details of $M_{\rm tag}$ signal regions in Ref.~\cite{zhangsf}. The $\mu^+$ selection is performed by using the $dE/dx$, TOF and EMC information and requires the muon hypothesis to be greater than the pion hypothesis. The inclusive MC yield is estimated by analyzing an inclusive MC sample corresponding to ten times luminosity of data.}
  \label{tab:CR}
  \centering
  \begin{tabular}{c|c|c|c|c}
    \hline\hline
    \diagbox{Requirement}{Process} & $D_s^+\to \mu^+\nu_\mu$ & other $\tau^+$ decay & $e^{+}e^{-}\to\tau^{+}\tau^{-}/q \bar{q}~(q = u, d, s)$ & $D_s^+\to \eta\pi^+$ \\ \hline
    $M_{\rm tag}$                    & Signal region & Signal region & Sideband region & Signal region \\
    $M_{\rm miss}^2$  (GeV$^2/c^4$)  & $\in(-0.2, 0.2)$ & $\in(0.6, 1.2)$ & $\in(0.6, 1.2)$ & $\in(0.2, 0.4)$ \\
    $E^\text{max}_\text{neu}$ (GeV)  & $< 0.3$ & $< 0.4$ & $> 0.1$ & $> 0.3$\\
    $\mu^+$ selection                & Yes & ... & ... & ... \\
    \hline\hline
    Data yield                       & 4725  & 8734  & 230984  & 4466  \\
    Inclusive MC yield               & 47180 & 83781 & 2162763 & 40428 \\
    Purity~($\%$)                    & 68.0  & 55.7  & 63.4    & 69.2  \\
    \hline\hline
  \end{tabular}
\end{table*}

Good data-MC consistencies can be seen in the comparisons of the BDT output scores for the four background sources bewteen data and MC simulation in different control regions.

\subsection{\boldmath Fit to data}
The signal yield of $D_s^+\to\tau^+(\to\pi^+\bar{\nu}_\tau)\nu_\tau$ is extracted from a maximum likelihood fit to the distribution of the BDT output scores for the data combined from all energy points. In the fit, the signal and background shapes are modeled with the simulated shapes derived from MC simulations and included as RooHistPdf objects~\cite{ROOT} in the fit, with both yields floated.

\begin{figure}[htbp]\centering
  \includegraphics[keepaspectratio=true,width=3.4in,angle=0]{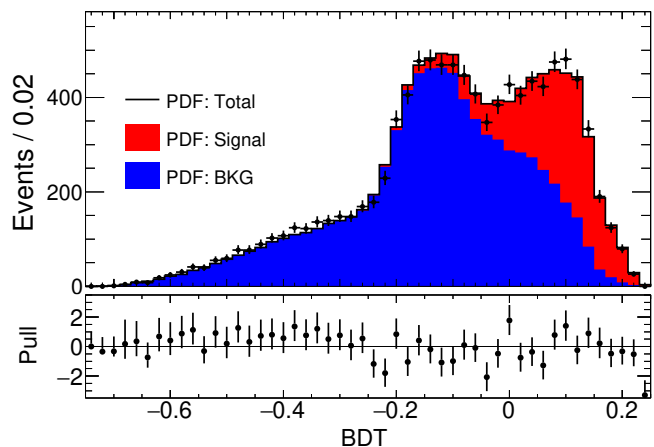}\\
  \caption{Fit result on the BDT score of the candidate events in data. The black points are data. The red solid-filled and blue solid-filled histograms represent the fitted signal and background shapes, respectively. The black-solid curve represents the total fit. The pull distribution of the fit result, derived with RooPullVar~\cite{ROOT}, is shown at the bottom.}
  \label{fig:bdtfit}
\end{figure}

Figure~\ref{fig:bdtfit} shows the result of the fit to the distribution of the BDT score in the signal region. From the fit, we obtain $2411\pm 75$ $D_s^+\to\tau^+(\to\pi^+\bar{\nu}_\tau)\nu_\tau$ events.

Our analysis procedure, including the extraction of ST and DT yields, has been validated by analyzing 40 full simulation samples separately, in which the average of the measured branching fractions is consistent with the branching fraction in the full simulation sample. Each of simulated sample has comparable luminosity of data. To further examine the stablility of the BDT fit method, we generate 10000 pseudo-data-sets with the observed data BDT distribution using the bootstrap method~\cite{bootstrap, ROOT}. We fit to these pseudo-data-sets individually. The pull of the fitted yield $p(N_\text{sig})$ is defined as
\begin{equation}\label{eq:pull}
  p(N_\text{sig}) = \frac{N^\text{pseudo}_\text{sig} - N^\text{real}_\text{sig}}{\sigma^\text{pseudo}_N},
\end{equation}
where $N^\text{real}_\text{sig}$ is the fitted yield of real data, $N^\text{pseudo}_\text{sig}$ and $\sigma^\text{pseudo}_N$ are the fitted yield and its statistical uncertainty of the pseudo-data-sets, respectively. The distribution of the $p(N_\text{sig})$ values is fitted with a Gaussian function. The mean and width values are obtained to be $-0.018 \pm 0.010$ and $1.010 \pm 0.007$, respectively.
These imply no bias of the BDT fit method.

\subsection{\boldmath Branching fraction result}

The DT efficiencies are shown in Table~\ref{tab:xyzDTefftau}. These efficiencies have been corrected by factors which takes into account the data-MC efficiency differences for the requirements of $\pi^+$ tracking and PID, $E^\text{max}_\text{neu},\,N_{\rm extra}^{\rm charge},\,N_{\rm extra}^{\pi^{0}}$, $E_{\gamma}$, EOP, $\rm{cos}\theta_{\rm{miss}}$ and $best~\gamma(\pi^0)$ selection as described in Sec.~\ref{sec:sys}. The individual correction factors are listed in Table~\ref{tab:corrfactor}.

The branching fraction of $D_s^+\to \tau^+\nu_\tau$ is determined to be $\b(D_s^+\to\tau^+\nu_\tau) = (5.44\pm0.17)\%$, where the uncertainty is statistical only and the branching fraction of $\tau^+\to \pi^+\bar \nu_\tau$ has been set to be 10.82\%~\cite{pdg2022}.

\begin{table*}[htb]\centering
  \caption{The $\text{DT}$ efficiencies ($\epsilon_\text{DT}$ in $\%$) of $D_s^+\to\tau^+(\to\pi^+\bar{\nu}_\tau)\nu_\tau$ for each tag mode and data sample. The uncertainties are statistical only. These efficiencies do not include the branching fractions of the intermediate decays of $K^0_S\to\pi^+\pi^-$, $\pi^0\to\gamma\gamma$, $\eta\to\gamma\gamma$, $\eta_{3\pi}\to\pi^+\pi^-\pi^0$, $\eta^\prime_{\pi\pi\eta}\to\pi^+\pi^-\eta$, $\eta^\prime_{\gamma\rho}\to\gamma\rho^0$ or $\rho\to\pi\pi$.}
  \label{tab:xyzDTefftau}
  \scalebox{1.00}
  {
    \begin{tabular}{c| c c c c c c c c}
      \hline\hline
      \diagbox{Tag mode}{$E_\text{cm}$ (GeV)} & $4.128$ & $4.157$ & $4.178$ & $4.189$ & $4.199$ & $4.209$ &  $4.219$ &  $4.226$ \\
      \hline
      {\ensurestackMath{
      \alignCenterstack{
      K^{+}K^{-}\pi^{-}\cr
      K^{+}K^{-}\pi^{-}\pi^{0}\cr
      \pi^{+}\pi^{-}\pi^{-}\cr
      K_{S}^{0}K^{-}\cr
      K_{S}^{0}K^{-}\pi^{0}\cr
      K^{-}\pi^{+}\pi^{-}\cr
      K_{S}^{0}K^{+}\pi^{-}\pi^{-}\cr
      K_{S}^{0}K^{-}\pi^{+}\pi^{-}\cr
      \pi^{-}\eta\cr
      \pi^{-}\eta^{\prime}_{\pi\pi\eta}\cr
      \pi^{-}\eta^{\prime}_{\gamma\rho}\cr
      \rho^{-}\eta\cr
      \rho^{-}\eta_{3\pi}}}} &
      {\ensurestackMath{
      \alignCenterstack{
      19.9\pm0.1\cr
      5.9\pm0.1\cr
      29.8\pm0.1\cr
      24.7\pm0.1\cr
      10.0\pm0.1\cr
      25.3\pm0.1\cr
      10.0\pm0.1\cr
      8.6\pm0.1\cr
      27.5\pm0.1\cr
      13.4\pm0.1\cr
      17.8\pm0.1\cr
      12.6\pm0.1\cr
      5.6\pm0.1}}} &
      {\ensurestackMath{
      \alignCenterstack{
      19.4\pm0.1\cr
      6.2\pm0.1\cr
      29.2\pm0.1\cr
      24.1\pm0.1\cr
      9.9\pm0.1\cr
      24.7\pm0.1\cr
      10.1\pm0.1\cr
      8.8\pm0.1\cr
      27.1\pm0.1\cr
      13.4\pm0.1\cr
      17.5\pm0.1\cr
      12.3\pm0.1\cr
      5.4\pm0.1}}} &
      {\ensurestackMath{
      \alignCenterstack{
      18.4\pm0.1\cr
      6.0\pm0.1\cr
      27.4\pm0.1\cr
      23.4\pm0.1\cr
      9.6\pm0.1\cr
      23.6\pm0.1\cr
      10.2\pm0.1\cr
      8.7\pm0.1\cr
      25.8\pm0.1\cr
      12.7\pm0.1\cr
      16.6\pm0.1\cr
      11.9\pm0.1\cr
      5.3\pm0.1}}} &
      {\ensurestackMath{
      \alignCenterstack{
      18.2\pm0.1\cr
      5.9\pm0.1\cr
      27.1\pm0.2\cr
      23.0\pm0.2\cr
      9.7\pm0.1\cr
      23.1\pm0.2\cr
      9.9\pm0.1\cr
      8.5\pm0.1\cr
      25.3\pm0.2\cr
      12.3\pm0.1\cr
      16.3\pm0.1\cr
      11.5\pm0.1\cr
      5.1\pm0.1}}} &
      {\ensurestackMath{
      \alignCenterstack{
      18.2\pm0.1\cr
      5.8\pm0.1\cr
      26.3\pm0.2\cr
      22.5\pm0.1\cr
      9.4\pm0.1\cr
      22.5\pm0.1\cr
      9.8\pm0.1\cr
      8.6\pm0.1\cr
      25.2\pm0.2\cr
      12.4\pm0.1\cr
      16.3\pm0.1\cr
      11.5\pm0.1\cr
      5.2\pm0.1}}} &
      {\ensurestackMath{
      \alignCenterstack{
      17.2\pm0.1\cr
      5.6\pm0.1\cr
      25.2\pm0.2\cr
      21.6\pm0.1\cr
      8.9\pm0.1\cr
      21.5\pm0.1\cr
      9.4\pm0.1\cr
      8.2\pm0.1\cr
      24.2\pm0.2\cr
      11.7\pm0.1\cr
      15.5\pm0.1\cr
      11.0\pm0.1\cr
      4.9\pm0.1}}} &
      {\ensurestackMath{
      \alignCenterstack{
      17.1\pm0.1\cr
      5.5\pm0.1\cr
      25.1\pm0.2\cr
      20.9\pm0.1\cr
      8.8\pm0.1\cr
      21.2\pm0.1\cr
      9.0\pm0.1\cr
      7.9\pm0.1\cr
      23.9\pm0.2\cr
      11.4\pm0.1\cr
      15.4\pm0.1\cr
      11.1\pm0.1\cr
      4.8\pm0.1}}} &
      {\ensurestackMath{
      \alignCenterstack{
      17.0\pm0.1\cr
      5.7\pm0.1\cr
      25.8\pm0.2\cr
      21.9\pm0.1\cr
      9.1\pm0.1\cr
      21.9\pm0.1\cr
      9.5\pm0.1\cr
      8.3\pm0.1\cr
      24.3\pm0.2\cr
      12.1\pm0.1\cr
      15.7\pm0.1\cr
      11.1\pm0.1\cr
      4.9\pm0.1}}} \\
      \hline\hline
    \end{tabular}
}
\end{table*}

\section{SYSTEMATIC UNCERTAINTIES}\label{sec:sys}
The sources of systematic uncertainties in the measurement of the branching fraction are divided into three categories. The first one is from the ST analysis procedure, the second one is from DT analysis procedure and the last one is from the fit to the BDT output score.

\subsection{\boldmath Uncertainties from ST analysis procedure}\label{sec:systNst}
The uncertainty of the fits to the $M_{\rm tag}$ spectra is estimated by varying the signal and background shapes and repeating the fit for both data and inclusive MC sample. The nominal signal shape is chosen as the one after requiring the angles between each reconstructed and generated track to be less than $15^\circ$. The alternative signal shape is obtained by varying the matching angle by $\pm 5^\circ$. The background shape is changed from nominal one to a third-order Chebychev polynomial. The relative change of the ST yields in data over the ST efficiencies is considered as the systematic uncertainty. Moreover, an additional uncertainty due to the background fluctuation of the fitted ST yields is included. The quadrature sum of these three terms, $0.52\%$, is assigned as the associated systematic uncertainty.

Due to different reconstruction environments in the inclusive and signal MC samples, the ST efficiencies determined by the inclusive MC sample may be different from those by the signal MC sample. This may lead to incomplete cancellation of the systematic uncertainties associated with the ST selection, referred to as ``tag bias". Inclusive and signal MC efficiencies are compared and the tracking and PID efficiencies for kaons and pions are studied for different track multiplicities. The resulting ST-average offsets are assigned as the systematic uncertainties from tag bias.

\subsection{\boldmath Uncertainties from DT analysis procedure}\label{sec:systeff}
The systematic uncertainties associated with DT event reconstruction and efficiency determination are considered as four parts: the tracks and neutrals reconstruction and identification, the signal MC sample sizes, the input branching fractions to the $\b(D_s^+\to\tau^+(\to\pi^+\bar{\nu}_\tau)\nu_\tau)$ determination and the basic event selections.

The systematic uncertainty in the $\gamma(\pi^0)$ selection is estimated by using a control sample of $J/\psi\to\pi^+\pi^-\pi^0$ decays~\cite{BES3GamSyst}, and the corresponding systematic uncertainty is assigned as $1\%$. The systematic uncertainties of $\pi^+$ tracking and PID are studied with control samples of light hadron processes produced in $e^+e^-$ collisions as summarized in Table~\ref{tab:corrfactor}. Small data-MC differences are found, as shown in Table~\ref{tab:corrfactor}. To compensate these differences, we correct the effective efficiency by these factors. After corrections, the residual statistical uncertainty is assigned as individual systematic uncertainty.

The uncertainty due to the limited sizes of the MC samples, which is used for the determination of the DT efficiencies, is $0.19\%$. The systematic uncertainties associated with the input branching fractions, $\b(D^{*+}_s\to\gamma(\pi^{0}) D^+_s)$ and $\b(\tau^+\to\pi^+\bar{\nu}_\tau)$, are examined by varying individual nominal values by $\pm 1\sigma$~\cite{pdg2022}. Combining these two effects in quadrature gives a total systematic uncertainty of $0.52\%$.

The systematic uncertainties of the requirements of $E^\text{max}_\text{neu},\,N_{\rm extra}^{\rm char},\,N_{\rm extra}^{\pi^0}$,~$E_{\gamma}$, $\text{EOP}$, $|\!\cos{\theta_\text{miss}}|$, and $best~\gamma(\pi^0)$ selection are studied with control samples of DT hadronic decays tagged by the same tag modes as in nominal analysis. The difference between data and simulation are corrected using factors from Table~\ref{tab:corrfactor} and the residual statistical uncertainties are assigned as systematic uncertainties for each source.

\begin{table*}[htb]\centering
  \caption{The used control samples and correction factors for different MC
  mismodelling sources. The uncertainties associated with correction factors arise from the statistical fluctuations in both the data and MC simulation.}
  \label{tab:corrfactor}
    \begin{tabular}{C{4.5cm} C{7.5cm} C{4.5cm}}
    \hline\hline
    Source & Control sample & Correction factor \\
    \hline
    $\pi^+$ tracking & $e^+e^-\to K^+K^-\pi^+\pi^-$ & $1.0033\pm0.0035$\\
    $\pi^+$ PID & $e^+e^-\to K^+K^-\pi^+\pi^-(\pi^0)$ and $\pi^+\pi^-\pi^+\pi^-(\pi^0)$ & $0.9890\pm0.0032$ \\
    $E^\text{max}_\text{neu},\,N_{\rm extra}^{\rm char},\,N_{\rm extra}^{\pi^0}$ & $D_s^+\to K^+K^-\pi^+$ and $D_s^+\to K^0_S K^+$  & $0.9918\pm0.0041$  \\
    $E_{\gamma}$ requirement & $D_s^+\to K^+K^-\pi^+$ and $D_s^+\to K^0_S K^+$ & $1.0066\pm0.0046$  \\
    $\text{EOP}$ requirement & $D_s^+\to K^+K^-\pi^+$ and $D_s^+\to K^0_S \pi^+$ & $0.9994\pm0.0014$  \\
    $\cos{\theta_\text{miss}}$ requirement & $D_s^+\to K^+K^-\pi^+$ & $1.0130\pm0.0083$  \\
    $Best~\gamma(\pi^0)$ selection & $D_s^+\to K^+K^-\pi^+$ and $D_s^+\to K^0_S K^+$  & $1.0035\pm0.0018$ \\
    \hline\hline
    \end{tabular}
\end{table*}

\subsection{\boldmath Uncertainty associated with BDT output score}\label{sec:systfit}
The systematic uncertainties associated with the fit to BDT output score are considered in three aspects. 

To examine the effect of the unobserved decays of $D_s^+\to \gamma \mu^+\nu_\mu$ and $D_s^+\to\pi^+\pi^0$, an alternative fit is performed, where these two decay components are added one by one. The yields of these decays are fixed to the corresponding experimental upper limits, $\b(D_s^+\to \gamma \mu^+\nu_\mu)<1.3\times10^{-4}$ and $\b(D_s^+\to\pi^+\pi^0)<3.4\times10^{-4}$~\cite{pdg2022}. Here we simply assume ${\mathcal B}(D_s^+\to \gamma\mu^+\nu_\mu)={\mathcal B}(D_s^+\to \gamma e^+\nu_e)$ based on lepton flavor universality. Eventually, their impact on $\b(D_s^+\to\tau^+\nu_\tau)$ is found to be negligible.

The branching fractions and the cross sections of the main background sources, as mentioned in Sec.~\ref{sec:furtherselection}, are varied within one standard deviation given in Ref.~\cite{pdg2022}. Quadratic sum of the relative changes of the re-measured branching fractions for each source, 1.50\%, is assigned as a systematic uncertainty.

Small data-MC differences in the input variables have been observed. To estimate their effect on the branching fraction measurement, we reweight all simulated variables to match individual data distributions iteratively. Quadratic sum of the relative changes of the fitted signal yield for each source, 0.69\%, is assigned as the systematic uncertainty.

By adding all systematic uncertainties in quadrature, as summarized in Table~\ref{tab:systsummary}, the total systematic uncertainty in the branching fraction measurement is determined to be $2.41\%$.

\begin{table}[htb]\centering
  \caption{Relative systematic uncertainties in the branching fraction measurement.}
  \label{tab:systsummary}
    \begin{tabular}{C{5.5cm} C{2.4cm}}
    \hline\hline
    Source &  Uncertainty~(\%) \\
    \hline
    ST yield  & $0.52$ \\
    Tag bias & $0.41$  \\
    \hline
    $\pi^+$ tracking & $0.35$\\
    $\pi^+$ PID & $0.32$ \\
    $\gamma(\pi^0)$ reconstruction & $1.00$ \\
    MC sample size  & $0.19$\\
    Input branching fractions   & $0.52$ \\
    Basic event selections & $1.06$ \\
    $M_{\rm miss}^{2}$ range & Negligible \\
    \hline
    $D_s^+\to\gamma\mu^+\nu_\mu$ background & Negligible \\
    $D_s^+\to\pi^+\pi^0$ background & Negligible \\
    Background estimate & $1.50$  \\
    Input shape for BDT & $0.69$ \\
    \hline
    Total & $2.41$ \\
    \hline\hline
    \end{tabular}
\end{table}

\section{RESULTS}\label{sec:result}
With the result for $\b(D_s^+\to\tau^+\nu_\tau)$ obtained in this study, we determine
\begin{equation}
  \nonumber
  \begin{array}{r c l c}
    f_{D_s^+}|V_{cs}| & = & (248.3\pm3.9_{\rm stat}\pm3.0_{\rm syst}\pm1.0_{\rm input})~\text{MeV}, &  \\
  \end{array}
\end{equation}
\noindent where the third uncertainty is from the external inputs of $m_\ell$, $m_{D_s^+}$, and $\tau_{D_s^+}$~\cite{pdg2022}.

By taking $|V_{cs}| = 0.97349\pm0.00016$ given by the SM~\cite{pdg2022} global fit as an input, we obtain
\begin{equation}
  \nonumber
  \begin{array}{r c l c}
    f_{D_s^+} & = & (255.0\pm4.0_{\rm stat}\pm3.1_{\rm syst}\pm1.0_{\rm input})~\text{MeV}, &  \\
  \end{array}
\end{equation}
which is in agreement with the LQCD predictions~\cite{flag2021}. Conversely, by taking the LQCD calculation of $f_{D_s^+} = 249.9\pm0.5$~MeV~\cite{flag2021} as an input, we determine
\begin{equation}
  \nonumber
  \begin{array}{r c l c}
    |V_{cs}| & = & 0.993\pm0.015_{\rm stat}\pm0.012_{\rm syst}\pm0.004_{\rm input}, & \\
  \end{array}
\end{equation}
which agrees well with the result given by the SM~\cite{pdg2022} global fit.

Using the method described in~\cite{averagemethod} which takes into account the correlation of systematic uncertainties, we obtain the average branching fraction to be $\b(D_s^+\to\tau^+\nu_\tau) = (5.33\pm0.07_{\rm stat}\pm0.08_{\rm syst})\%$ by combining the BESIII measurements of the branching fractions of $D_s^+\to\tau^+\nu_\tau$ measured via $\tau^+\to\pi^+\pi^0\bar{\nu}_\tau$~\cite{BES3taupipi0}, $\tau^+\to e^+\bar{\nu}_\tau\nu_e$~\cite{BES3taue}, $\tau^+\to \mu^+\bar{\nu}_\tau\nu_\mu$~\cite{BES3taumu}, and that via $\tau^+\to\pi^+\bar{\nu}_\tau$ from this study. Here the uncertainties from the ST yield, the $\pi^+$ tracking and PID, the soft $\gamma(\pi^0)$ reconstruction, the $best~\gamma(\pi^0)$ selection, and the tag bias are taken to be correlated. Additional common uncertainties come from $\tau_{D_s^+}$, $m_{D_s^+}$ and $m_\tau$ for $f_{D_s^+}$ and $|V_{cs}|$, while all the other uncertainties are independent. This gives $f_{D_s^+} = (252.4\pm1.7_{\rm stat}\pm1.8_{\rm syst}\pm1.0_{\rm input})~\text{MeV}$ and $|V_{cs}| = 0.983\pm0.007_{\rm stat}\pm0.007_{\rm syst}\pm0.004_{\rm input}$. Combining the world average of $\b(D^+_s\to \mu^+\nu_\mu) = (5.43\pm0.15)\text{\textperthousand}$~\cite{pdg2022}, we obtain $R = \Gamma(D_s^+\to\tau^+\nu_\tau)/\Gamma(D_s^+\to\mu^+\nu_\mu) = 9.81\pm0.33$.

Averaging the branching fractions of $D_s^+\to\tau^+\nu_\tau$ measured by CLEO~\cite{CLEOel,CLEOpi,CLEOro}, BaBar~\cite{BaBarlnu}, Belle~\cite{Bellelnu}, BESIII~\cite{BES3taupipi0,BES3taumu,BES3taue,BES34009} and from this study, we obtain the average branching fraction to be $\b(D_s^+\to\tau^+\nu_\tau) = (5.37\pm0.10)\%$. This gives $f_{D_s^+} = (253.3\pm2.3_{\rm stat, syst}\pm1.0_{\rm input})~\text{MeV}$, $|V_{cs}| = 0.987\pm0.009_{\rm stat, syst}\pm0.004_{\rm input}$ and $R = \Gamma(D_s^+\to\tau^+\nu_\tau)/\Gamma(D_s^+\to\mu^+\nu_\mu) = 9.89\pm0.33$.

\begin{table*}[htb]\centering
  \caption{Comparison of the branching fractions and the corresponding products of $f_{D_s^+}|V_{cs}|$ from various experiments. ``Weighted'' are obtained by combining with considering the correlated effects. ``Average'' are obtained by combining both statistical and systematic uncertainties, but not the third uncertainties, which are dominated by the uncertainty of the $D_s^+$ lifetime. The uncertainty of ``Average'' of $\b$ and the first uncertainty of ``Average" of $f_{D^+_s}|V_{cs}|$ are
  the combined values of their statistical and systematic uncertainties, respectively, and the second uncertainty of ``Average" of $f_{D^+_s}|V_{cs}|$ due to the uncertainty of the quoted $D_s^+$ lifetime.}
  \label{tab:resultsummary}
  \def\1#1#2{\multicolumn{#1}{#2}}
    \begin{tabular}{L{2.0cm} C{2.0cm} C{2.5cm} C{3cm} C{3.8cm} C{3.5cm}}
    \hline\hline
    Experiment & $E_{\rm cm}$ (GeV) & Mode & $\tau^+$ decay & $\b~(\%)$ & $f_{D_s^+}|V_{cs}|$ (MeV)\\
    \hline
    {\bf This work}   & {\bf 4.128-4.226} & $\pmb{D^\pm_sD^{*\mp}_s}$  & $\pmb{\pi^+\bar{\nu}_\tau}$  & $\pmb{5.44\pm0.17\pm0.13}$ & $\pmb{248.3\pm3.9\pm3.1\pm1.0}$ \\
    BESIII~\cite{BES3taumu} & 4.128-4.226 &  $D^\pm_sD^{*\mp}_s$ & $\mu^+\bar{\nu}_\tau\nu_\mu$  & $5.37\pm0.17\pm0.15$ & $246.7\pm3.9\pm3.6\pm1.0$ \\
    BESIII~\cite{BES3taue} & 4.178-4.226 & $D^\pm_sD^{*\mp}_s$ & $e^+\bar{\nu}_\tau\nu_e$  & $5.27\pm0.10\pm0.13$ & $244.4\pm2.3\pm2.9\pm1.0$ \\
    BESIII~\cite{BES3taupipi0} & 4.178-4.226 &  $D^\pm_sD^{*\mp}_s$ & $\pi^+\pi^0\bar{\nu}_\tau$  & $5.30\pm0.25\pm0.20$ & $245.1\pm5.8\pm4.7\pm1.0$ \\
    BESIII~\cite{hajime} & 4.178-4.226 &  $D^\pm_sD^{*\mp}_s$  & $\pi^+\bar{\nu}_\tau$  & $5.21\pm0.25\pm0.17$ & $243.0\pm5.8\pm4.0\pm1.0$ \\
    \hline
    Weighted$^{\text{a}}$   & $\cdot\cdot\cdot$ &  $\cdot\cdot\cdot$  & $\cdot\cdot\cdot$   & $5.33\pm0.07\pm0.08$  & $245.7\pm1.7\pm1.8\pm1.0$ \\
    \hline
    BESIII~\cite{BES34009} & 4.008 &  $D_s^+D_s^-$  & $\pi^+\bar{\nu}_\tau$   & $3.28\pm1.83\pm0.37$ & $192.8\pm44.2\pm10.9\pm0.8$ \\
    CLEO~\cite{CLEOel}   & 4.170 & $D^\pm_sD^{*\mp}_s$   & $e^+\bar{\nu}_\tau\nu_e$ & $5.30\pm0.47\pm0.22$ & $245.1\pm10.9\pm5.1\pm1.0$ \\
    CLEO~\cite{CLEOpi}  & 4.170 & $D^\pm_sD^{*\mp}_s$   & $\pi^+\bar{\nu}_\tau$   & $6.42\pm0.81\pm0.18$ & $269.7\pm17.2\pm3.8\pm1.1$ \\
    CLEO~\cite{CLEOro}  & 4.170 &  $D^\pm_sD^{*\mp}_s$  & $\rho^+\bar{\nu}_\tau$ & $5.52\pm0.57\pm0.21$ & $250.1\pm13.0\pm4.8\pm1.0$ \\
    BaBar~\cite{BaBarlnu}  & 10.56 &  $DKX\gamma D^{-}_s$ & $e^+\bar{\nu}_\tau\nu_{e}, \mu^+\bar{\nu}_\tau\nu_{\mu}$ & $4.96\pm0.37\pm0.57$ & $237.1\pm8.9\pm13.7\pm1.0$ \\
    Belle~\cite{Bellelnu}  & 10.56 &  $DKX\gamma D^{-}_s$   & $\pi^+\bar{\nu}_\tau, e^+\bar{\nu}_\tau\nu_{e}, \mu^+\bar{\nu}_\tau\nu_{\mu}$ & $5.70\pm0.21^{+0.31}_{-0.30}$ & $254.1\pm4.7\pm7.0\pm1.0$\\
    \hline
    Average   &  $\cdot\cdot\cdot$ & $\cdot\cdot\cdot$   & $\cdot\cdot\cdot$   & $5.37\pm0.10$  & $246.6\pm2.2\pm1.0$ \\
    \hline\hline
    \1{4}{l}{$^{\text{a}}$It excludes ``BESIII~\cite{hajime}''.} \\
  \end{tabular}
\end{table*}

\section{SUMMARY}\label{sec:summary}
Using 7.33 fb$^{-1}$ of $e^+e^-$ collision data taken at $E_{\rm cm}$ between 4.128 and 4.226 GeV, we report the updated study of $D_s^+\to\tau^+\nu_\tau$ via $\tau^+\to \pi^+\bar{\nu}_\tau$, where the candidates are maximally separated from the background distribution using a BDT. The branching fraction of $D_s^+\to \tau^+\nu_\tau$ is determined to be $(5.44\pm0.17_{\rm stat}\pm0.13_{\rm syst})\%$. This result is consistent with the previous measurements by CLEO~\cite{CLEOel,CLEOpi,CLEOro}, BaBar~\cite{BaBarlnu}, Belle~\cite{Bellelnu}, and BESIII~\cite{BES3taupipi0,BES3taumu,BES3taue}. In particular, it supersedes the previous BESIII result of $(5.21\pm0.25_{\rm stat}\pm0.17_{\rm syst})\%$ published in Ref.~\cite{hajime}, which was measured via $\tau^+\to \pi^+\bar{\nu}_\tau$ in a narrower $M^2_{\rm miss}$ range by analyzing 6.32 fb$^{-1}$ of $e^+e^-$ collision data taken at $E_{\rm cm}$ between 4.178 GeV and 4.226 GeV. Table~\ref{tab:resultsummary} shows comparison of $\b(D^+_s\to\tau^+\nu_\tau)$ and $f_{D_s^+}|V_{cs}|$ obtained in this study and the previous measurements.

\acknowledgments
The BESIII Collaboration thanks the staff of BEPCII and the IHEP computing center for their strong support. This work is supported in part by National Key R\&D Program of China under Contracts Nos. 2020YFA0406400, 2020YFA0406300; National Natural Science Foundation of China (NSFC) under Contracts Nos. 11635010, 11735014, 11835012, 11935015, 11935016, 11935018, 11961141012, 12022510, 12025502, 12035009, 12035013, 12061131003, 12192260, 12192261, 12192262, 12192263, 12192264, 12192265; the Chinese Academy of Sciences (CAS) Large-Scale Scientific Facility Program; the CAS Center for Excellence in Particle Physics (CCEPP); Joint Large-Scale Scientific Facility Funds of the NSFC and CAS under Contract No. U1832207; CAS Key Research Program of Frontier Sciences under Contracts Nos. QYZDJ-SSW-SLH003, QYZDJ-SSW-SLH040; 100 Talents Program of CAS; The Institute of Nuclear and Particle Physics (INPAC) and Shanghai Key Laboratory for Particle Physics and Cosmology; ERC under Contract No. 758462; European Union's Horizon 2020 research and innovation programme under Marie Sklodowska-Curie grant agreement under Contract No. 894790; German Research Foundation DFG under Contracts Nos. 443159800, 455635585, Collaborative Research Center CRC 1044, FOR5327, GRK 2149; Istituto Nazionale di Fisica Nucleare, Italy; Ministry of Development of Turkey under Contract No. DPT2006K-120470; National Research Foundation of Korea under Contract No. NRF-2022R1A2C1092335; National Science and Technology fund of Mongolia; National Science Research and Innovation Fund (NSRF) via the Program Management Unit for Human Resources \& Institutional Development, Research and Innovation of Thailand under Contract No. B16F640076; Polish National Science Centre under Contract No. 2019/35/O/ST2/02907; The Royal Society, UK under Contract No. DH160214; The Swedish Research Council; U. S. Department of Energy under Contract No. DE-FG02-05ER41374.

\end{document}